\documentclass[12pt]{article}
\usepackage[left=1in,right=1in,top=1in,bottom=1in]{geometry}
\usepackage{amsfonts}
\usepackage{amsmath}
\usepackage{natbib}
\bibpunct{(}{)}{;}{{\it a}}{}{,}
\usepackage{amssymb}%
\usepackage{graphicx}
\usepackage{float}
\newcommand{\qed}{\nobreak \ifvmode \relax \else
   \ifdim\lastskip<1.5em \hskip-\lastskip
   \hskip1.5em plus0em minus0.5em \fi \nobreak
   \vrule height0.75em width0.5em depth0.25em\fi}

\begin{document}
\baselineskip=17pt

\title{The Area and Population of Cities: New Insights from a Different Perspective on Cities
\footnote{Rozenfeld and Makse: City College
of New York (email: hernanrozenfeld@gmail.com and hmakse@lev.ccny.cuny.edu). Rybski: Potsdam Institute for Climate Impact Research (email: diego.rybski@pik-potsdam.de). Gabaix: New York University, CEPR and NBER (email: xgabaix@stern.nyu.edu). This work is supported by the NSF through grant
SES-0624116 and DMS-0527518. We thank
L.H. Dobkins and J. Eeckhout for providing the data on MSA and M. Batty for providing data on GB and useful discussions; C.
Briscoe and R. Tumarkin for help with the manuscript; and S.
Brakman, M. Davis, G. Duranton, H. Garretsen, E. Rossi-Hansberg, Y. Ioannides, P. Krugman,
C. van Marrewijk, P.-D. Sarte and seminar participants at NYU,
Princeton and the Richmond Fed for helpful comments. }}


\author{Hern\'an D. Rozenfeld, Diego Rybski, \\
Xavier Gabaix, Hern\'an A. Makse}




\date{\today}

\medskip
\maketitle



The distribution of the population of cities has attracted a great
deal of attention, in part because it sharply constrains models of
local growth. However, to this day, there is no consensus on the
distribution below the very upper tail, because available data need
to rely on the ``legal'' rather than ``economic'' definition of
cities for medium and small cities. To remedy this difficulty, in
this work we construct cities ``from the bottom up'' by clustering
populated areas obtained from high-resolution data. This method
allows us to investigate the population and area of cities for urban
agglomerations of all sizes. We find that Zipf's law (a power law
with exponent close to 1) for population holds for cities as small
as 12,000 inhabitants in the USA and 5,000 inhabitants in Great
Britain. In addition the
distribution of city areas is also close to a Zipf's law. We provide
a parsimonious model with endogenous city area that is consistent
with those findings. (\textit{JEL} D30, D51, J61, R12)

\clearpage

\section{Introduction}
\label{intro}


This paper builds on a recently-proposed algorithm to construct cities
based on geographical features of high-quality micro data~\citep{RozenfeldPNAS}, rather
than informative but somewhat arbitrary legal or administrative definitions. It allows
us to take a fresh look at key quantities in urban economics, namely
the population and the area of cities.
We find that Zipf's law for population holds quite well, and well
below the very upper tail of the city size distribution, where it
had been shown to hold to a good degree of
approximation~\citep{GabaixIoannides04}. We also find that the distribution of city areas follows a power law, with
an exponent close to 1, the Zipf value. These findings help
constrain further theories of cities and theories of geography. We
present a baseline parsimonious model of cities, which features
endogenous city area, and is consistent with these two key stylized
facts, as well as others.

A key difficulty in studying cities is finding a practical way to
define them
\citep{Zipf49,Krugman96, Eaton97,Dobins00,Eeckhout04,Soo05,batty}.
A canonical method
involves defining Metropolitan Statistical Areas (MSAs)
obtained in the USA from the US Census Bureau~\citep*{MSA}. MSAs are
defined for each major agglomeration, and attempt to capture their
extent
by merging administratively defined entities, counties in the
USA, based on their social or economic ties. For instance, the MSA of
Boston includes not only the administrative unit of Boston, but also
adjacent Cambridge, MA.
MSAs derive their appeal from a strong economic logic, but their
construction requires qualitative analysis and is very
time-consuming. Therefore,
MSAs have been constructed only for the 276 most populated cities in
the USA, and the corresponding Zipf's law has been documented only
for the upper tail of the
distribution~\citep{GabaixIoannides04,Soo05}.

Two main alternatives to the MSAs have been proposed in the
literature. One method is to use administrative or legal borders of
cities to define the so-called ``places'' as done
by~\citet{Eeckhout04} and \citet{Levy09}. The analysis of 25,359
places in the USA has suggested that Zipf's law holds in the upper
tail~\citep{Levy09} but
fails in the bulk of the distribution, as legally defined cities
follow a log-normal distribution
rather than a power-law~\citep{Eeckhout04,eeckhout09}.
The advantage of this definition is that it allows the study of the distribution of cities of all sizes.
Still, it is problematic to define cities through their fairly
arbitrary legal boundaries (the places method treats Cambridge and
Boston as two separate units), and indeed, this is why researchers
prefer agglomerations such as MSAs whenever such constructs are
available. A second approach is to construct cities from micro data
\citep{Holmes08,duranton05,mori08,michaels08}. In particular,
\citet{Holmes08} consider cities to be individual cells of
six-by-six miles,
for which the tail of the city size distribution
is much less fat-tailed than Zipf's law. However,
this is probably because constraining cities to areas of six-by-six
miles makes it nearly impossible to find a very large city. Hence,
because of these methodological difficulties, the shape of distribution of
agglomerations beneath the few hundred largest cities is still an open problem.

Here we build on an algorithm, the
City Clustering Algorithm (CCA), that  was recently introduced in \citep{RozenfeldPNAS} and based on previous studies done by~\citet{Makse95} to build cities ``from the
bottom-up''.
The algorithm
defines a \textquotedblleft city\textquotedblright \ as a maximally
connected cluster of populated sites defined at high resolution.
Namely, a population cluster is made of contiguous populated sites
within a prescribed distance $\ell$ that cannot be expanded: all
sites immediately outside the cluster have a population density
below a cutoff threshold. Rather than defining a city as one cell,
as done by~\citet{Holmes08}, our method defines an agglomeration as
a maximally connected cluster of potentially many cells.

We find that Zipf's law holds, to a good approximation, in the USA
and GB, for both populations and areas. We also find that density
has only a weak correlation with population and area. We propose
that the two facts of Zipf's law for populations and areas could
serve as tight constraints on models of cities.
As we can measure area, we wish to model it. Hence, we provide a
parsimonious urban model that incorporates areas, and generates Zipf's law
for areas and populations.

In Section~\ref{dataandmethods} we present the analyzed data and
explain the CCA. In Section~\ref{population} we present our results
for the population distribution of CCA clusters in the USA and GB.
We also compare the CCA clusters with US Census MSAs and places and
present a formal test of robustness of our clustering method. In
Section~\ref{area} we show the results of the area distribution of
CCA clusters in the USA and GB and present a study of the
correlations between densities, areas, and populations for CCA
clusters. In Section ~\ref{model} we propose a model that can
integrate the findings, and we summarize our conclusions in Section~\ref{conclusions}.


\section{Data and Methods}
\label{dataandmethods}

\subsection{Raw data}

\begin{figure}[h]
\begin{centering}
\resizebox{0.6\textwidth}{!}{\includegraphics{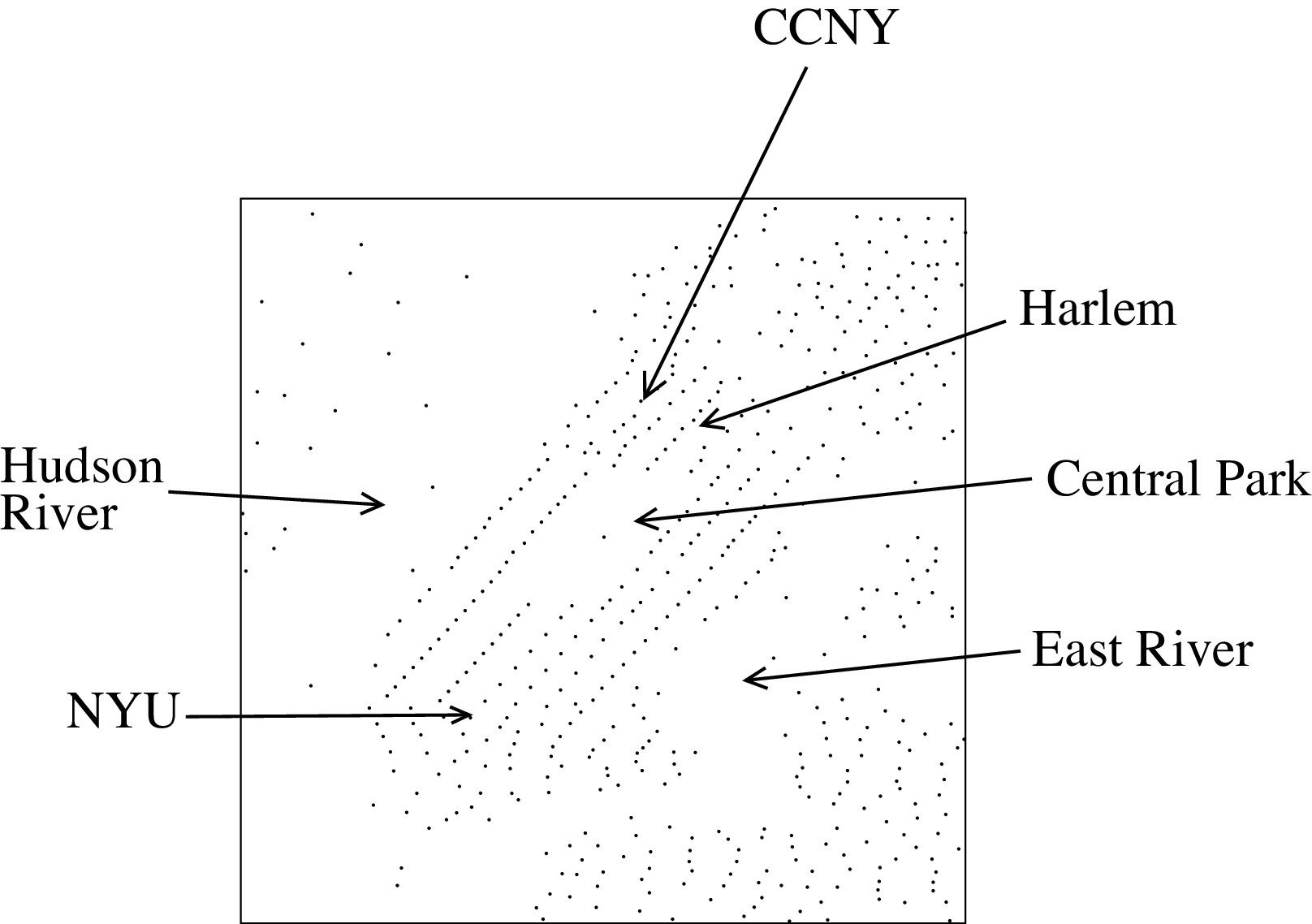}}
\caption{{\bf Raw data for Manhattan.} In this plot we show all FIPS codes
corresponding to Manhattan Island obtained  from the raw data
for the USA. Each point corresponds to a FIPS code specified by the
US Census Bureau.} \label{fips_central_park}
\end{centering}
\end{figure}

The data for the USA consists of the location and population of
61,224 points located throughout the area of the
USA~\citep*{USbureau}. Each point corresponds to a Federal
Information Processing Standard (FIPS) census tract code~\citep*{fips} generated by the US Census Bureau ranging in
population from 1,500 to 8,000 people, with a typical size of about 4,000 people.
FIPS codes are
uniquely specified by 11 digits. The first 2 digits correspond to the
state code, the next 3 to the county within the state, and the next 6
correspond to the census tract code. For example FIPS 36061016500
corresponds to New York State (36), New York County (Manhattan, 061),
census tract 016500 which is an area ranging from 58th Street to 60th
Street and from 8th Ave. to 9th Ave.  Figure~\ref{fips_central_park}
shows all FIPS for Manhattan Island in New York City and its
surroundings. The location of the FIPS is not always equidistant. For instance the shortest distance between 
two FIPS is about 100 m as in appears in Manhattan, while in less populated areas like Wyoming, FIPS can be separated by about 100 km.

The data for Great Britain (GB) is uniformly gridded at high resolution. It consists of
a grid with cell size $200$ m overlaid on the area of GB for which
the population in each cell is given.
The source of the GB data is the ESRC~\citep*{UKCensus} and is
composed of 5.75 million square cells comprising a total population
of about 55 million inhabitants in 1991.
Given that the GB data is more fine-grained that of the US, it is
arguably higher-quality.
All datasets and results used and presented in this work may be downloaded from our web page.

\subsection{The City Clustering Algorithm (CCA)}

\begin{figure}[!h]
\begin{centering}
{\bf a}\resizebox{0.4\textwidth}{!}{\includegraphics{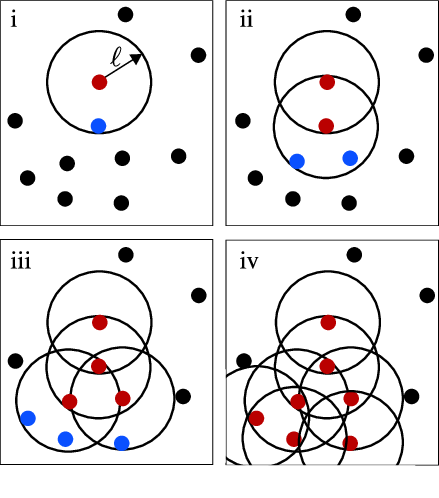}}
{\bf b}\resizebox{0.44\textwidth}{!}{\includegraphics{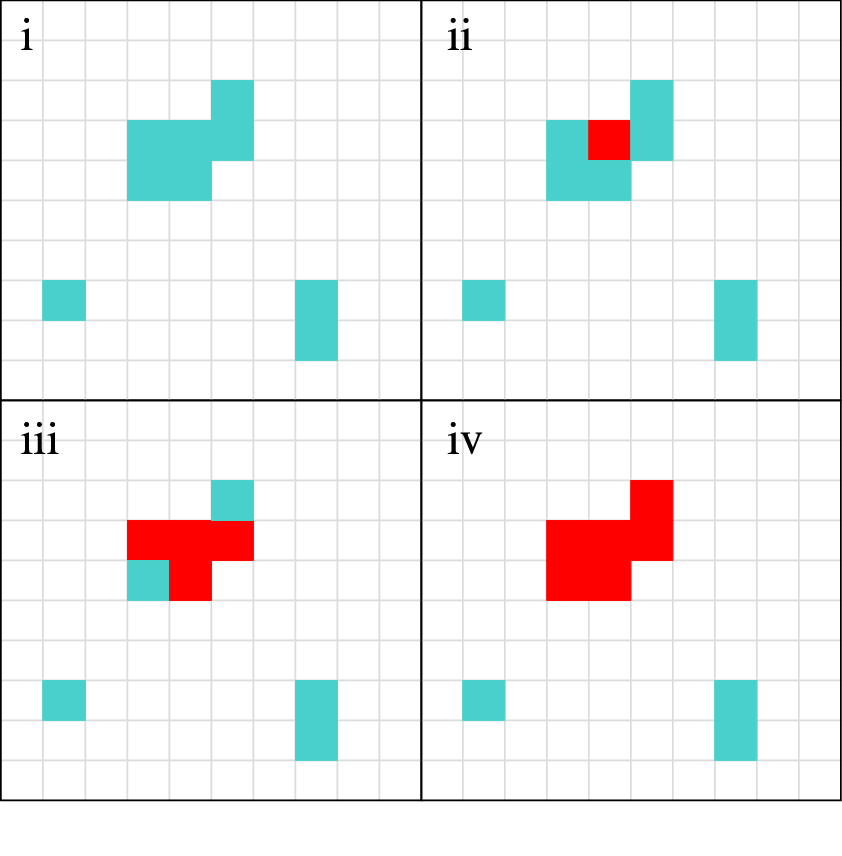}}
\caption{{\bf a,} CCA applied to the USA (continuum CCA). The points in this figure denote populated sites or FIPS. For our studies (and in this diagram) we use a density threshold $D_*=0$. {\bf (i)} We start
a cluster selecting a populated site, red point, among all available populated sites. We draw a circle of radius $\ell$ and add all
populated sites, blue point, that fall within the circle. {\bf (ii)} We draw a circle from the new member of this cluster and add all
populated sites (denoted by the two blue points) within the circle. {\bf (iii)} Recursively, we keep drawing circles from all new cluster
sites. The populated sites inside the circles (three blue points in this case) are merged into the cluster. {\bf (iv)} The red points are
the members of the cluster. Since no black point is at a distance smaller than $\ell$ from any red point, the cluster does not grow
anymore. We start the process again selecting another initial point that has not been already assigned to any cluster. This
process is repeated until all populated sites are assigned to a cluster. Notice that the choice of the initial condition, the first
selected point, does not influence the outcome.
{\bf b,} CCA applied to GB (discrete CCA). {\bf (i)} Cells are colored in blue if they are populated,
otherwise they are in white. {\bf (ii)} We
initialize the CCA by selecting a random populated cell (red cell). 
Then, we merge all populated neighbors of the
red cell as shown in {\bf (iii)}. We keep growing the cluster
by iteratively merging neighbors of the red cells until all
neighboring cells are unpopulated, as shown in {\bf (iv)}. 
Next, we pick another unburned populated cell and repeat the
algorithm until all populated cells are assigned to a cluster.
}
\label{CCA_details}
\end{centering}
\end{figure}

We start this section by providing a detailed explanation of the
CCA~\citep{RozenfeldPNAS}. In Fig.~\ref{CCA_details}a we show four steps
of the CCA when it is applied to the USA. To define a CCA cluster,
we first locate a populated site. Then, we
recursively grow the cluster by adding all nearest-neighbor sites
(populated sites within a distance smaller than the coarse-graining
level, $\ell$, from any site within the cluster) with a population
density, $D$, larger than a threshold $D_*$. The cluster stops
growing when no site outside the cluster with population density
$D>D_*$ is at a distance smaller than $\ell$ from the cluster boundary.
In this work, to minimize the number of free
parameters, we set the threshold $D_*= 0$, and therefore clusters are
recursively grown by merging all populated sites within a distance
smaller than $\ell$ from any site within
the cluster.

\begin{figure}
\begin{centering}
{\bf a} \includegraphics[width=0.6\textwidth]{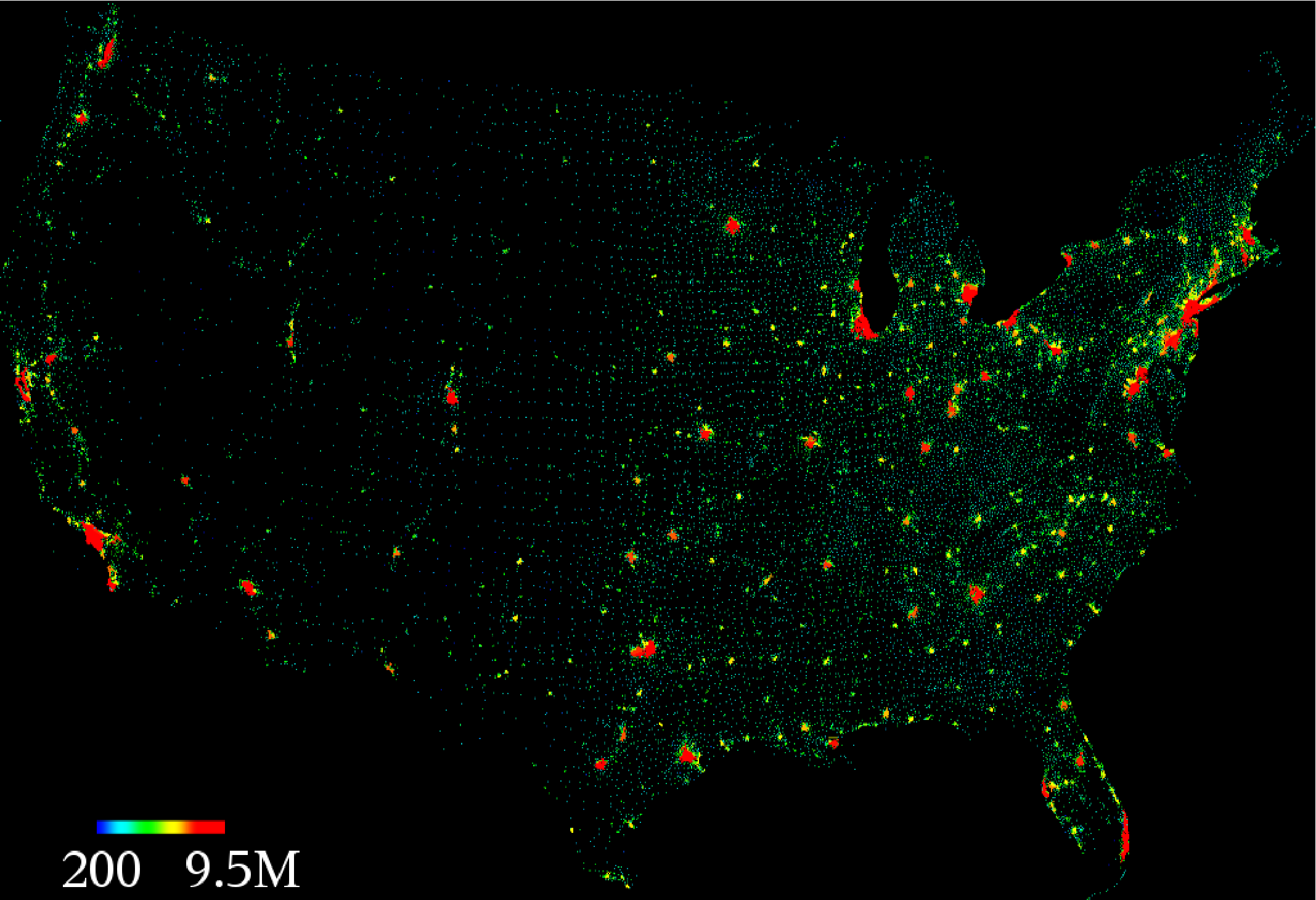}
{\bf b} \includegraphics[width=0.6\textwidth]{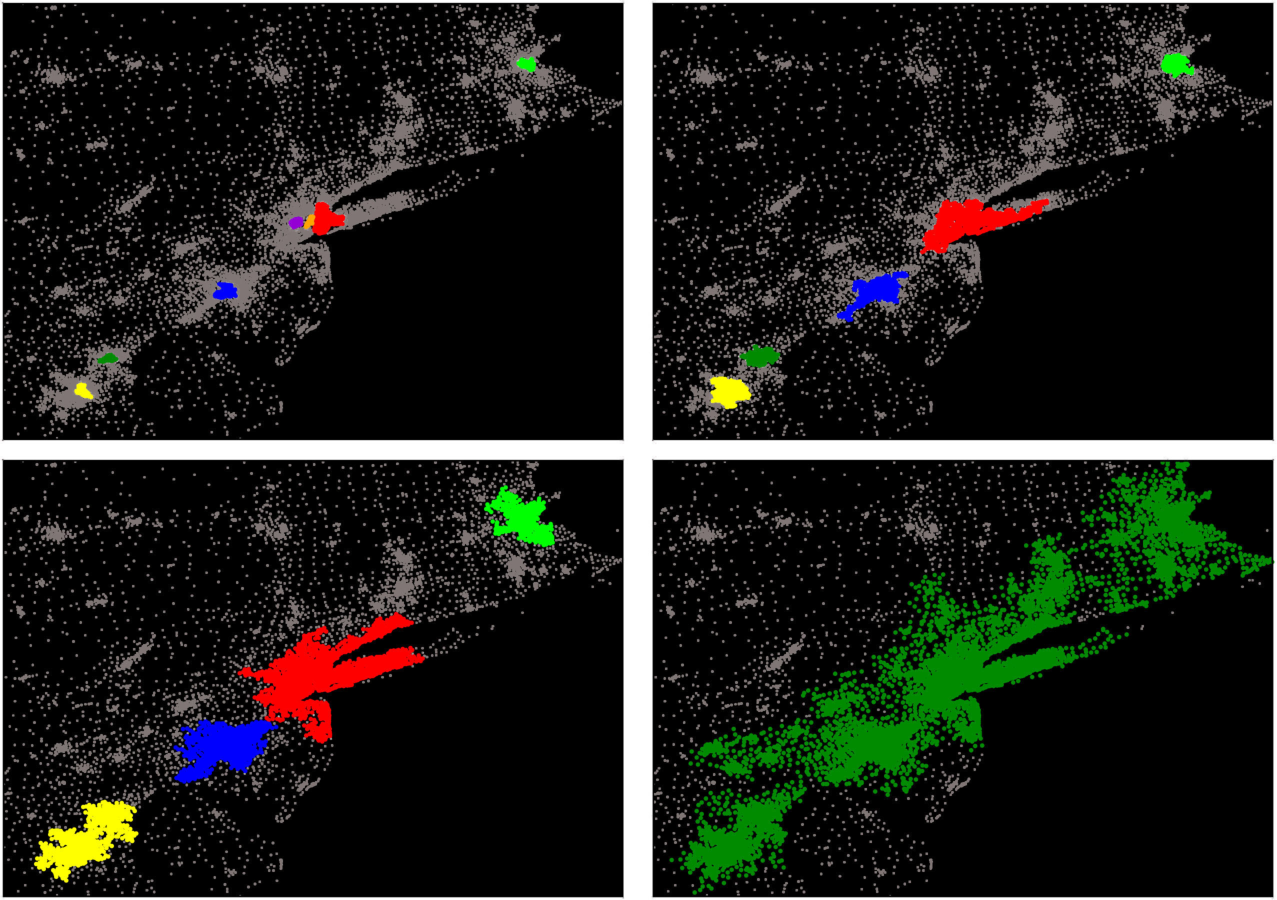}
\caption{{\bf CCA clusters in the USA. a,} CCA clusters applied to the entire USA. The map shows the
different clusters obtained by the algorithm.  The color indicates the population of each urban cluster (in
logarithmic scale).
{\bf b,} Results of the CCA applied to the major clusters of the northeastern
USA at different length scales. The top left panel shows the CCA
clusters for $\ell=$ 1 km separating the cities of
Washington D.C., Baltimore, Philadelphia, Newark, Jersey City, New York, and Boston.
The top right panel shows the results of the
algorithm when the data is coarse-grained to $\ell = 2$
km. Here, for example, the cities of New York, Newark and Jersey
City become part of the same cluster. The lower left panel shows the
results for $\ell=4$ km, where the main clusters are Washington D.C.-Baltimore; Philadelphia; New York-Newark-Jersey
City-Long Island; and Boston-Cambridge. The lower right panel for $\ell = 8$
km shows a giant cluster comprising all major cities in the northeastern USA. The gray points are also identified as part of other
clusters
but for clarity we do not specify them with individual colors in this figure.}
\label{usa_map}
\end{centering}
\end{figure}

Once the clusters are built, we calculate the population of a cluster as the sum of the
populations of all sites within the cluster.
Figure~\ref{usa_map}a shows a map of all identified clusters in
the continental USA where colors correspond to the cluster population, and Fig.~\ref{usa_map}b shows a detail of the clusters in the northeastern USA for different $\ell$.

Since  the data of GB is already gridded, the CCA algorithm adopts a simpler form that in the USA. To apply the CCA to the GB data we start from a populated cell and at each step we grow the cluster by adding all populated cells
neighboring the boundary of the cluster (see Fig~\ref{CCA_details}b). The cluster stops growing when all cells neighboring the cluster have a density no greater than $D_*$. 

As mentioned before, the data for GB differs from that of the USA, consisting of a grid with cell size
$200$ m overlaid on the map of GB. For this reason, since the data is already gridded at a
very high resolution, we simply merge cells from the original data to
obtain larger levels of coarse-graining at different grid sizes
$\ell$. We call this version of the CCA, the discrete CCA, while the version applied to the USA is the continuum CCA (see Fig.~\ref{CCA_details}).

\section{Population Distribution}
\label{population}

\subsection{Basic Results}
\label{basicresults}

\begin{figure}
\centering {
\includegraphics[width=0.6\textwidth]{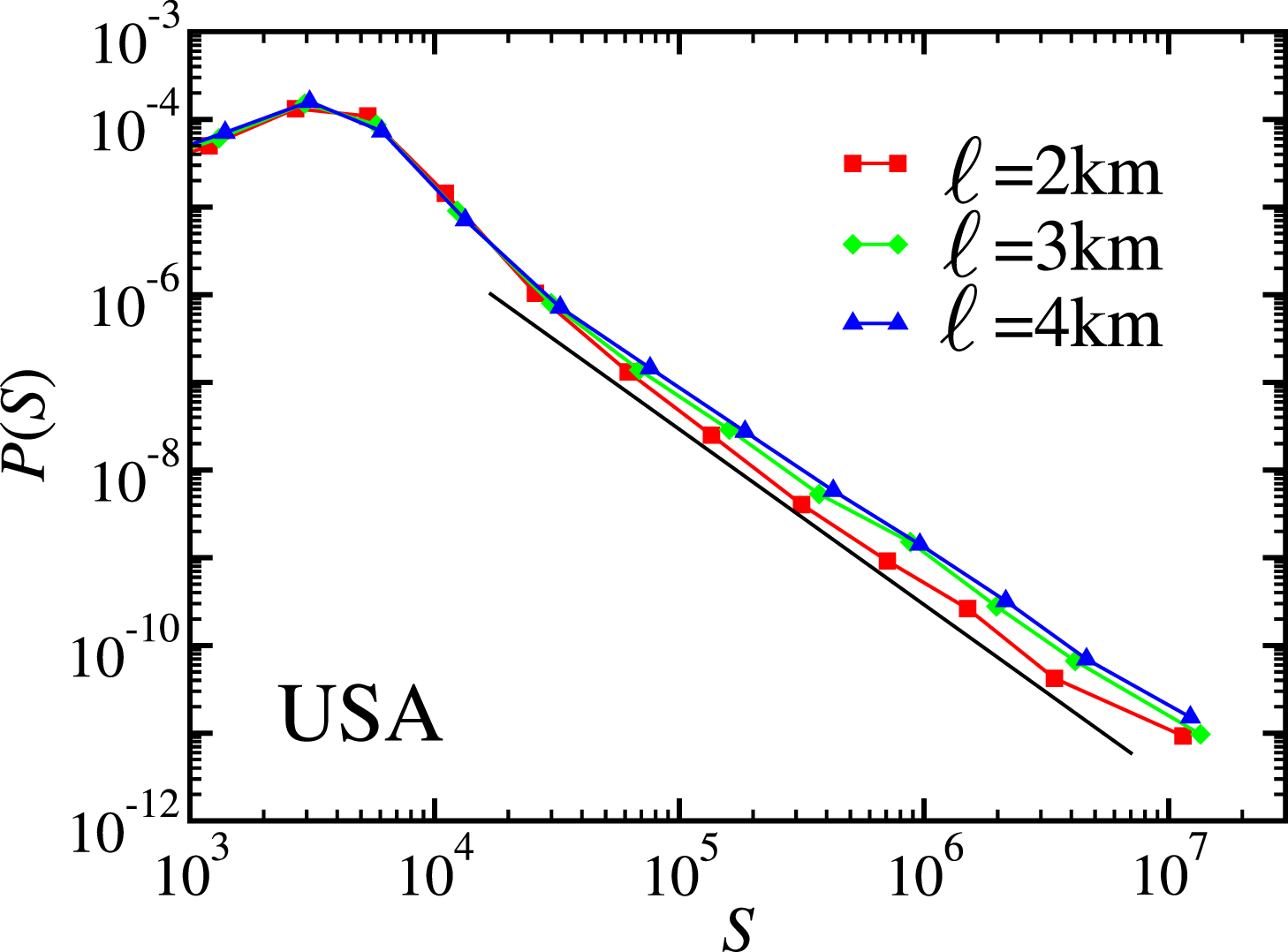}
}
\caption{Probability distribution of cluster populations $P(S)$
for the USA at different coarse-graining scales $\ell$. The
black solid line denotes a power-law function with exponent -2,
i.e. Zipf's Law.}
\label{ps}
\end{figure}
\begin{figure}
\centering {
\includegraphics[width=.6\textwidth]{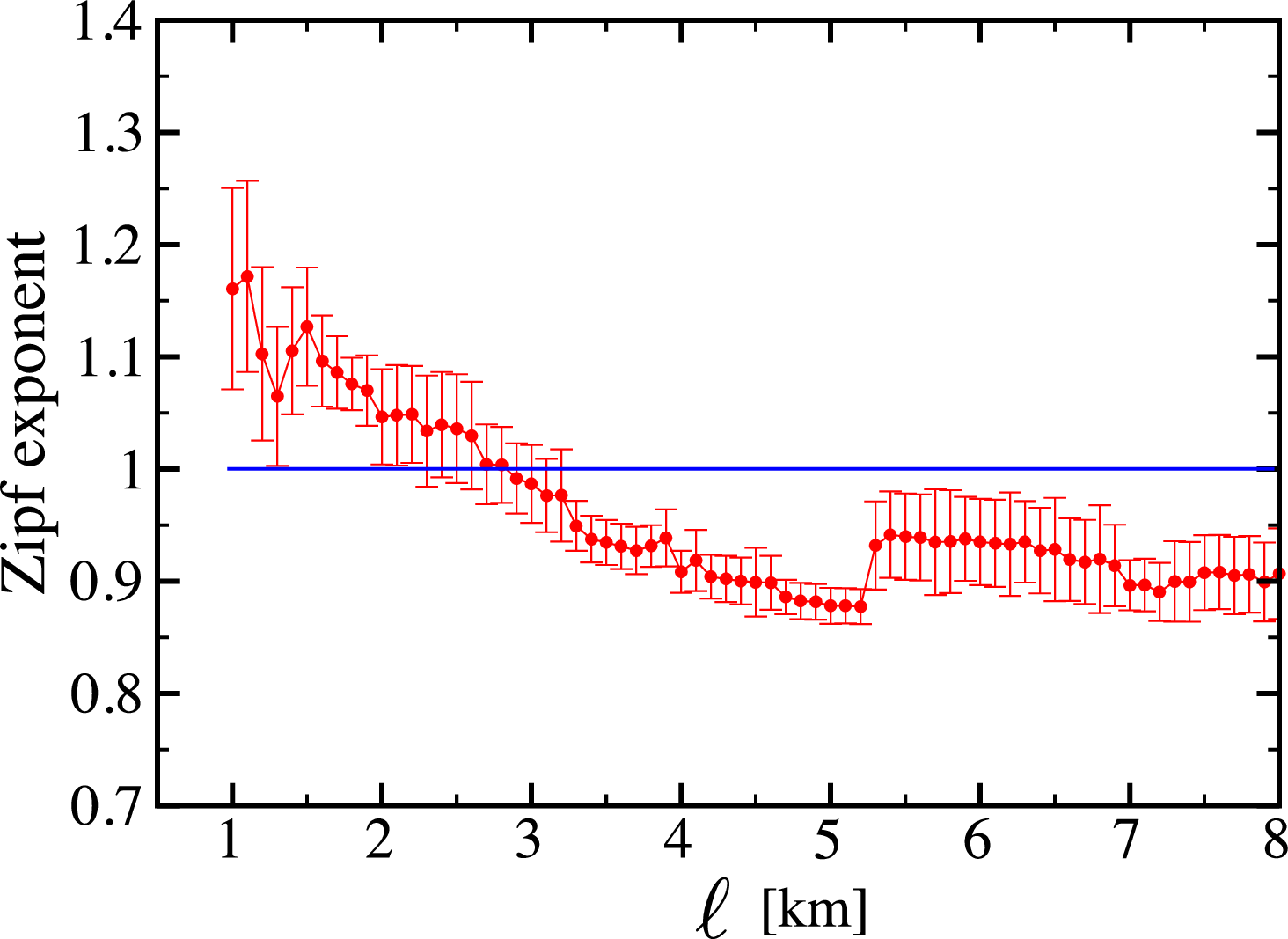}
}
\caption{Zipf exponent $\zeta$ obtained for
the USA clusters at different $\ell$. The error bars correspond to
$\pm 1$ standard deviation.}
\label{zeta_exp}
\end{figure}

We analyze the population data in the USA and GB to
obtain the distribution, $P(S)$, measuring the probability density
that a cluster has a population between $S$ and $S+{\rm
d}S$. Figure~\ref{ps} shows the results of $P(S)$ for the USA for
$\ell=2$ km, $\ell=3$ km, and $\ell=4$ km for which we obtain 30,201,
23,499, and 19,912 clusters, respectively. We find that the population
distribution follows a power-law of the form:
\begin{equation}
\label{zipf}
P \left(S\right)  \sim S^{-\zeta-1},
\end{equation}
with an exponent of $\zeta \approx 1$, in approximate accordance
with the value of Zipf's law. For example, when we estimate the
exponent for $\ell=3$ km and for clusters with $S > S_* = 12,000$
inhabitants (comprising 63\% of the country's population) we find
$\zeta= 0.97 \pm 0.03$ using an OLS estimator (the notation $\pm$
means that the standard deviation is 0.03). Figure~\ref{zeta_exp}
shows the Zipf exponent $\zeta$ for the USA for several value of
$\ell$. We observe that the exponent $\zeta$ remains approximately
within 5\% of the Zipf value in the range $\ell \in [2.5,3.5]$
km.

\begin{figure}
\centering {
\includegraphics[width=.6\textwidth]{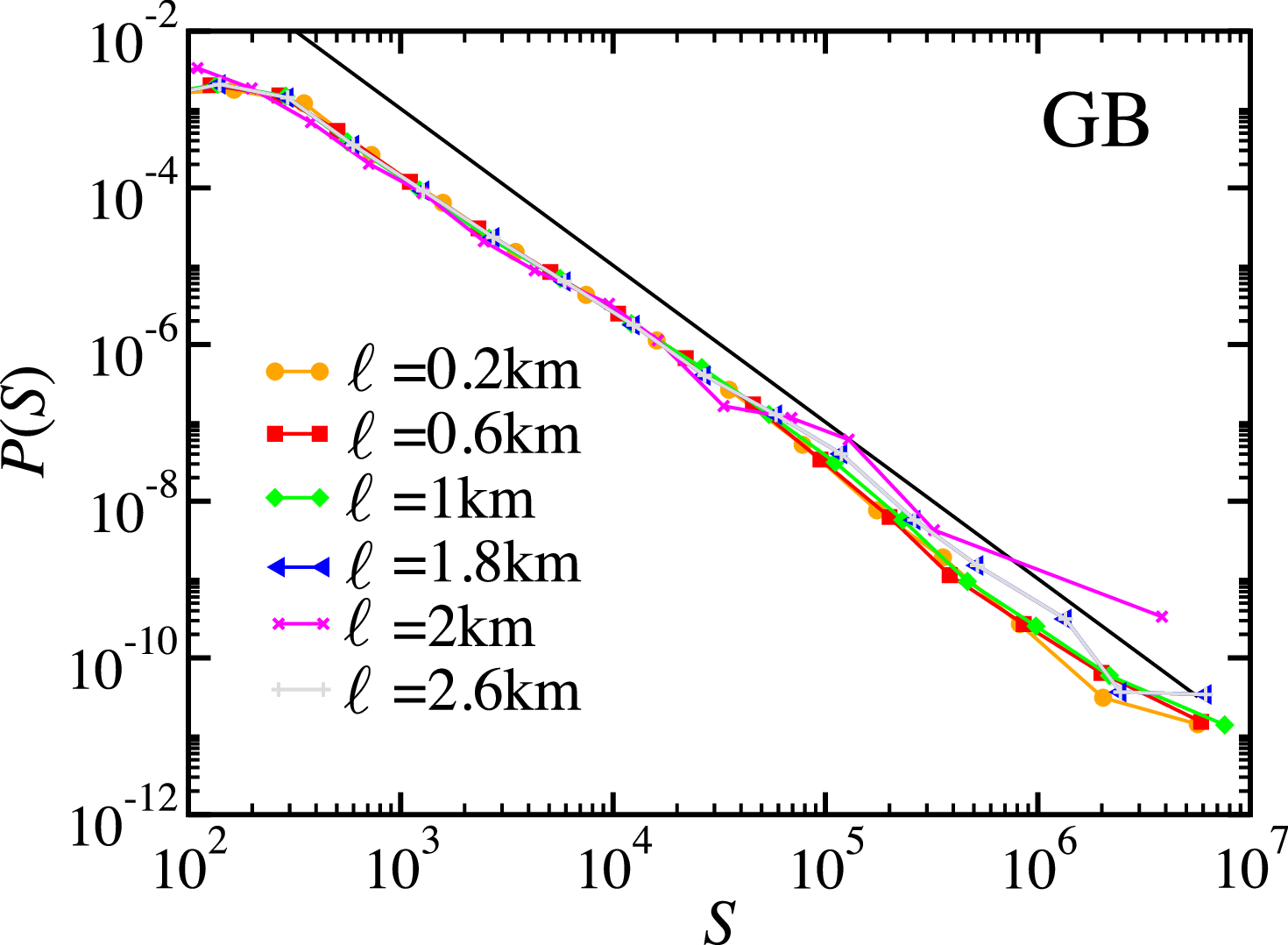}
}
\caption{Probability distribution $P(S)$ for
GB clusters at different coarse-graining scales $\ell$.
}
\label{uk}
\end{figure}

Figure~\ref{uk} displays the
population distribution of the CCA clusters in GB for $\ell=0.2$ km, $\ell=0.6$ km, $\ell=1$ km, $\ell=1.8$ km, $\ell=2$ km, and $
\ell=2.6$ km. For
clusters with a population above a cutoff $S_{\ast}=5,000$
inhabitants, the GB population follows a power-law to a good degree of
approximation. Using an OLS regression, we estimate for $\ell=1$ km
(1,008 clusters with 83\% of the
country's population) a Zipf exponent $\zeta = 1.07 \pm
0.03$. As in the case of the
USA, the exponent is similar for different choices of grid size
$\ell$.

To formally study the validity of our power-law fits, we employ the
test proposed by~\citet{GabaixIbragimov10} and \citet{Gabaix09}, offering a simple
quantification of possible deviations from a pure power-law. The
test for quadratic deviations is used to determine
if a power-law is adequate to describe the city size distribution.
The method is as follows. Sort the cities according to their rank
$i$ ($i=1$ being the largest city) and run the OLS regression
\begin{equation}
{\rm ln}(i-1/2) = {\rm constant} - \zeta~{\rm ln} S_i + q~({\rm
ln}S_i - \gamma)^2
\end{equation}
where $\zeta$ (the power law exponent) and $q$ (the quadratic
deviation from a power law) are the parameters to estimate and
$\gamma \equiv \left({\rm cov}(({\rm ln} S_j)^2,{\rm ln} S_j)\right)/\left(2{\rm
var}({\rm ln}S_j)\right)$. The recentering term $\gamma$ ensures that the
exponent $\zeta$ is the same whether the quadratic term is included
or not, and therefore $\zeta$ may be estimated beforehand using a
simple linear OLS. The quadratic test formalizes the intuition that
a pure power law has $q=0$ in the asymptotic limit, so a high value
of $|q|$ indicates deviations from power-law behavior. Under the null of a power law,
 for large samples $\sqrt{2N}q_N/\zeta^2$ converges to a standard normal distribution
 (where $N$ is the number of data-points). With probability 0.99, 
 a standard normal is less than 2.57 in absolute value. Hence, let $q_c \equiv
2.57 \zeta^2 / \sqrt{2N}$, be the critical value for the absolute value of the
quadratic term $q$ at the 1\% confidence level. If $|q|>q_c$ we reject
the hypothesis that the data is well described by a power-law since
the quadratic term becomes significant. Otherwise, if $|q|<q_c$, the
quadratic term is insignificant and we do not reject the power-law
hypothesis.

For the USA, when we consider the distribution of city sizes for
cities larger than $S_*=12,000$ for $\ell=3$ km, we obtain $|q| =
0.0291$ and $q_c = 0.0413$. Since $|q|<q_c$, we conclude that we can
disregard the quadratic correction to the OLS fit and consider that
the power-law describes the empirical distribution of city sizes. In
the case of GB, we consider $S_*=5,000$ and $ \ell=1$ km, for which
$|q|=0.0521$ and $q_c=0.0522$. Although $|q|$ and $q_c$ are very
close, the fact that $|q|<q_c$ indicates that we cannot reject the
hypothesis that the power-law describes the city size distribution
for GB.
We conclude that Zipf's law is a good description of city sizes with
population above $S_{\ast}=12,000$ inhabitants in the USA and $S_{\ast}=5,000$ inhabitants in GB. This
comprises $1,947$ clusters (for $\ell=3$ km) and a population of
171.3 million out of a total population of 271.1 million in the USA, and 1,007 clusters (for $\ell=1$ km) and a population of 45.3 milllion out of a total population of 54.5 million in GB, in
contrast to previous samples
\citep{Soo05} typically having a few
hundred cities.

So far, we have only focused on the part of the distribution where a power law fit could not be statistically rejected. Now, somewhat more loosely, we turn to a visual inspection of Fig.~\ref{ps} and Fig.~\ref{uk}. We see that the distribution is arguably well-approximated by a power law, in a region covering cities above 3,000 inhabitants in the USA, and cities above 300 inhabitants in GB. The deviations from the power law, while statistically significant, are not very large economically.  Hence, we also submit that, for the modelling a cities, the domain of an approximate power law is quite large. This domain comprises 17609 clusters and a population of 259.3 million (96\% of the total population) in the USA, and 9214 clusters and a population of 53.1 million (96\% of the total population) in GB.


\subsection{Comparison between CCA clusters, MSAs, and Places}

\begin{figure}
\centering {
\hbox{{\bf a}
\resizebox{0.5\textwidth}{!}{\includegraphics{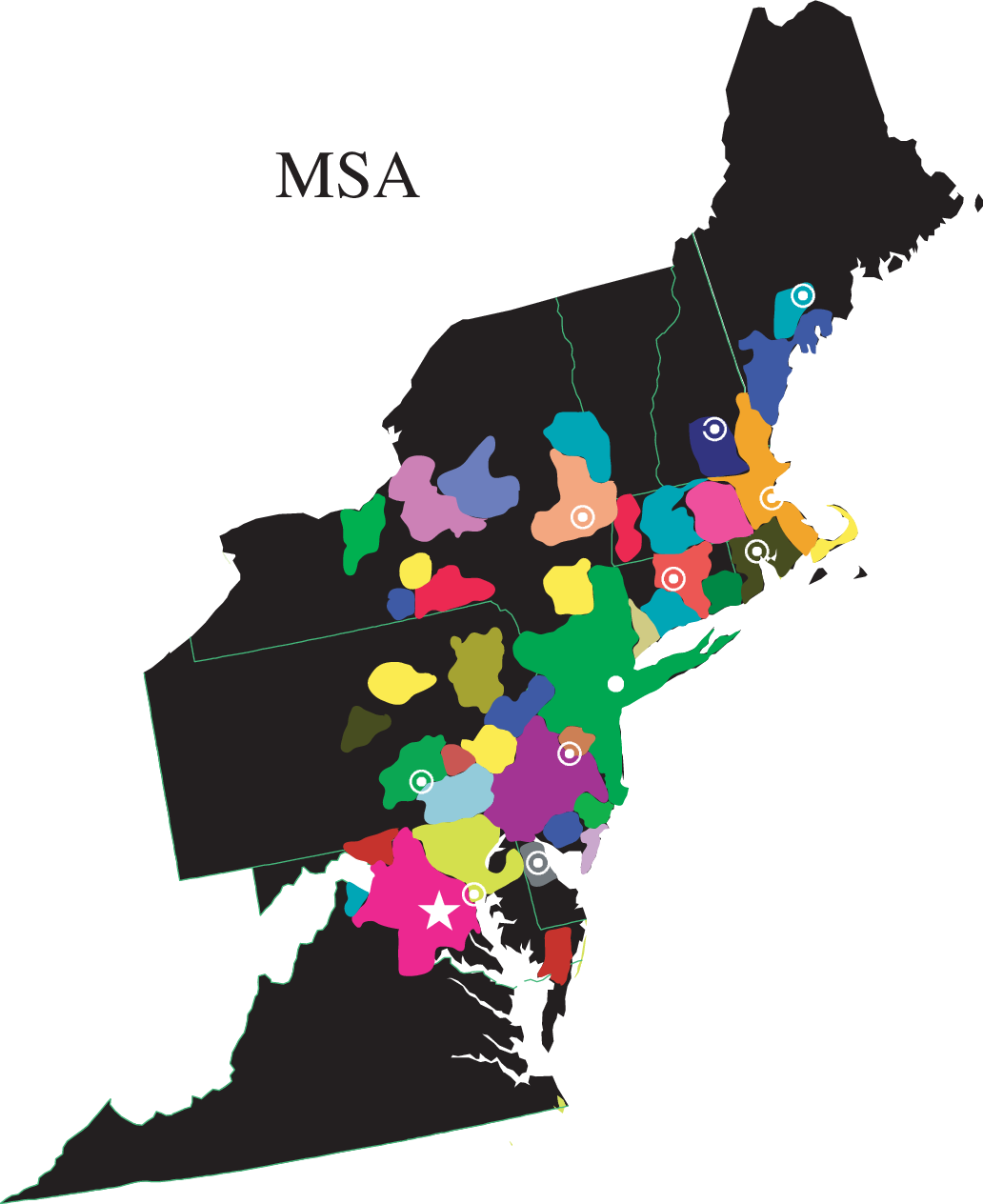}}
{\bf b} \resizebox{0.5\textwidth}{!}{\includegraphics{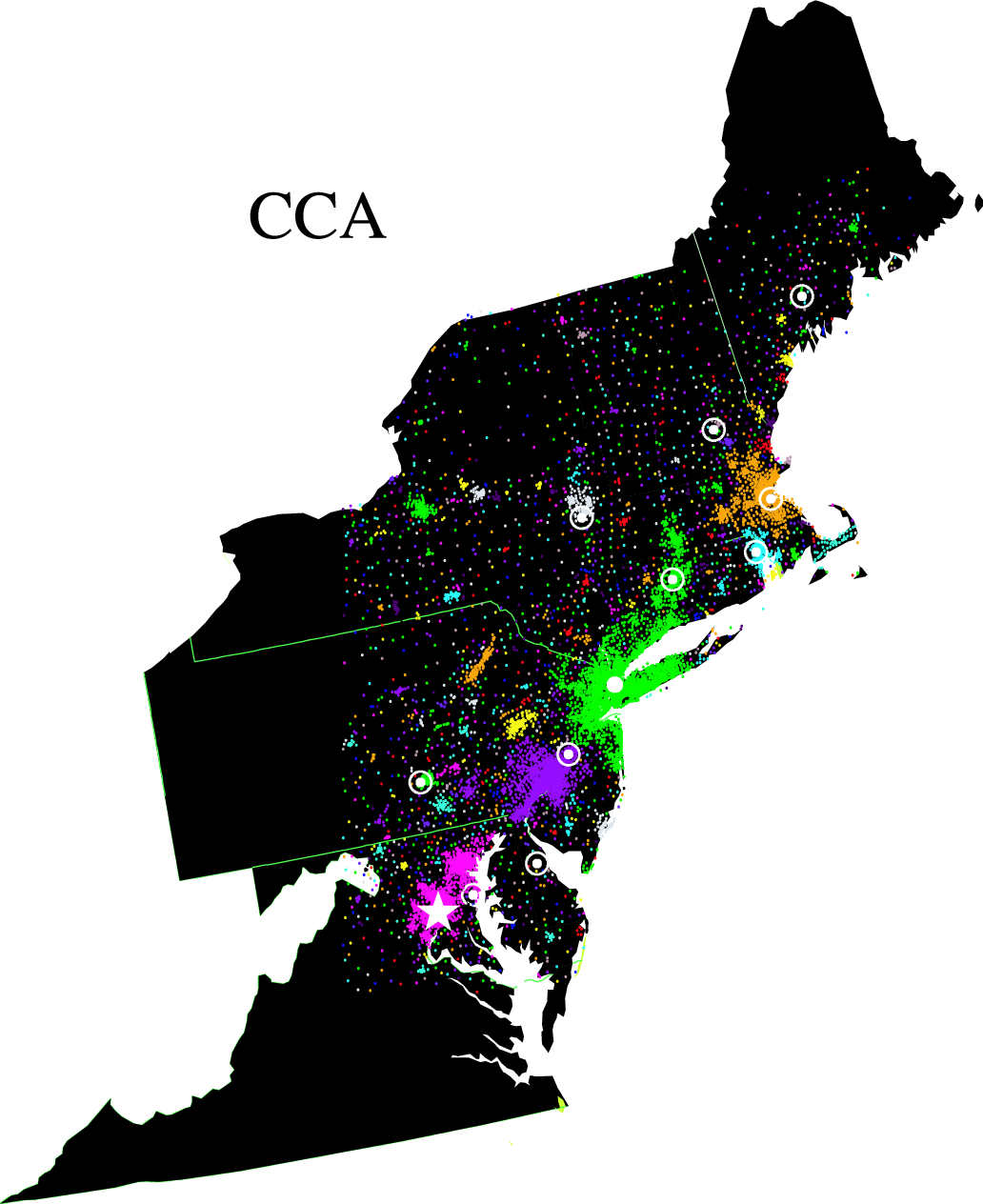}}
}}
\caption{{\bf Comparison between the MSAs and the CCA clusters.} {\bf
a,} MSAs for the northeastern USA.  For example, New York county
(Manhattan) with a population larger than 50,000 is a center of a
MSA.  Jersey City belongs to the same MSA since a large number of
its population commute to Manhattan, setting economic and social
ties between the two regions.  {\bf b,} CCA clusters for the
northeastern USA for $\ell=5$ km. Each cluster or MSA is plotted
with a different color.  For instance, the MSA centered in New York
City (in green in {\bf a}) is composed of several clusters. The
largest overlapping cluster found with the CCA is in green in {\bf
b}.  The white concentric circles correspond to the location of
the state capitals in the considered region. The star denotes
Washington D.C. and the white full circle corresponds to New York
City.}
\label{overlap}
\end{figure}

Although the CCA allows one to choose the observation level of
population clusters, $\ell$, it may be desirable to have an objective
way to choose $\ell$. For this purpose, we perform a comparison with
the MSAs in the USA which may be considered a benchmark for plausibly
well-constructed cities.  MSAs are defined starting from a highly
populated central county with population larger than 50,000 and
adding its surrounding counties if they have social or economic ties
such as large commuting patterns between the regions.
Figures ~\ref{overlap}a and ~\ref{overlap}b show a comparison between
the MSAs of the northeastern USA and the clusters obtained using CCA.

\begin{figure}
\centering { \hbox{ 
{\bf a} \includegraphics[width=0.45\textwidth]{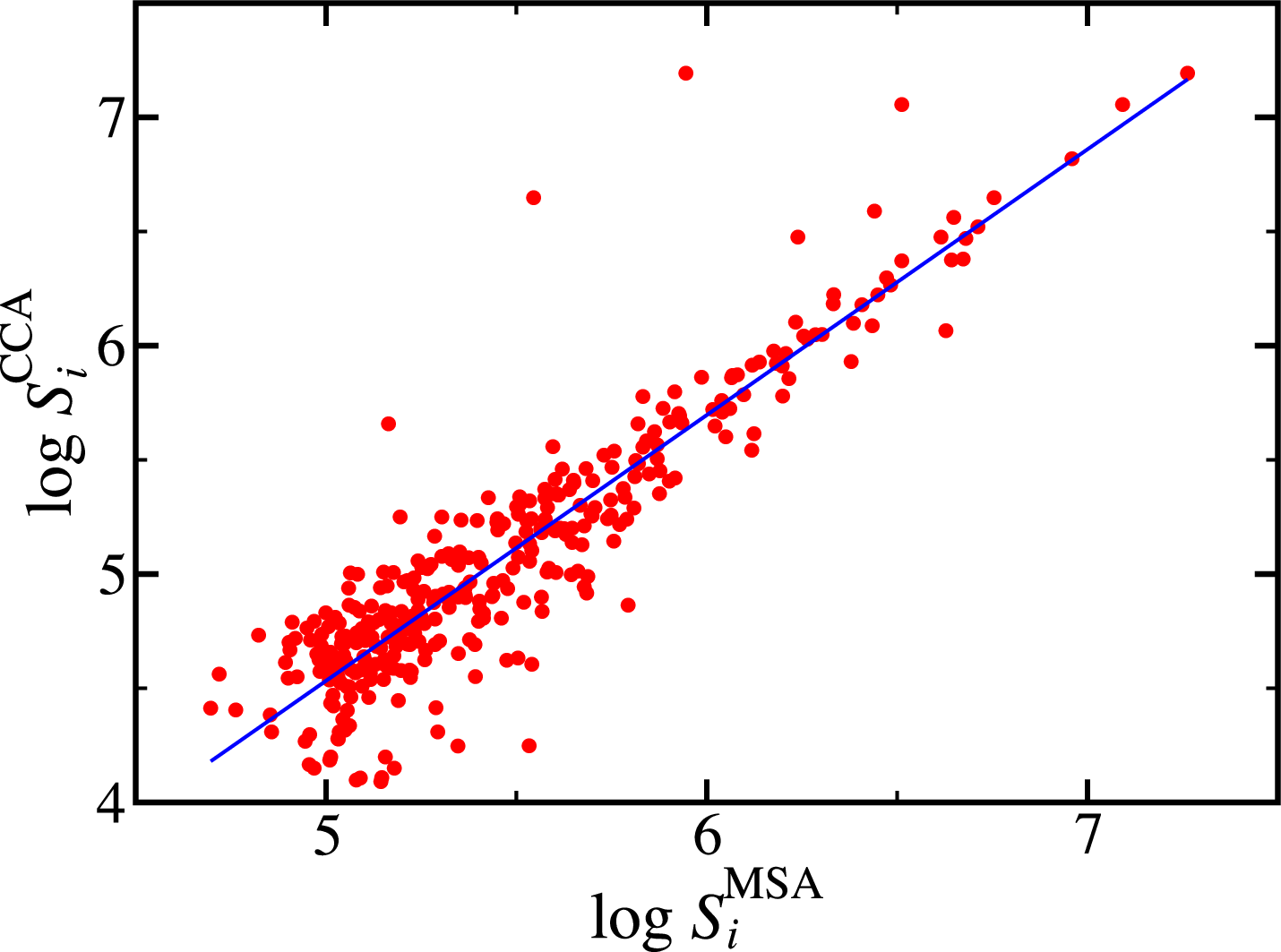}
{\bf b} \includegraphics[width=.44\textwidth]{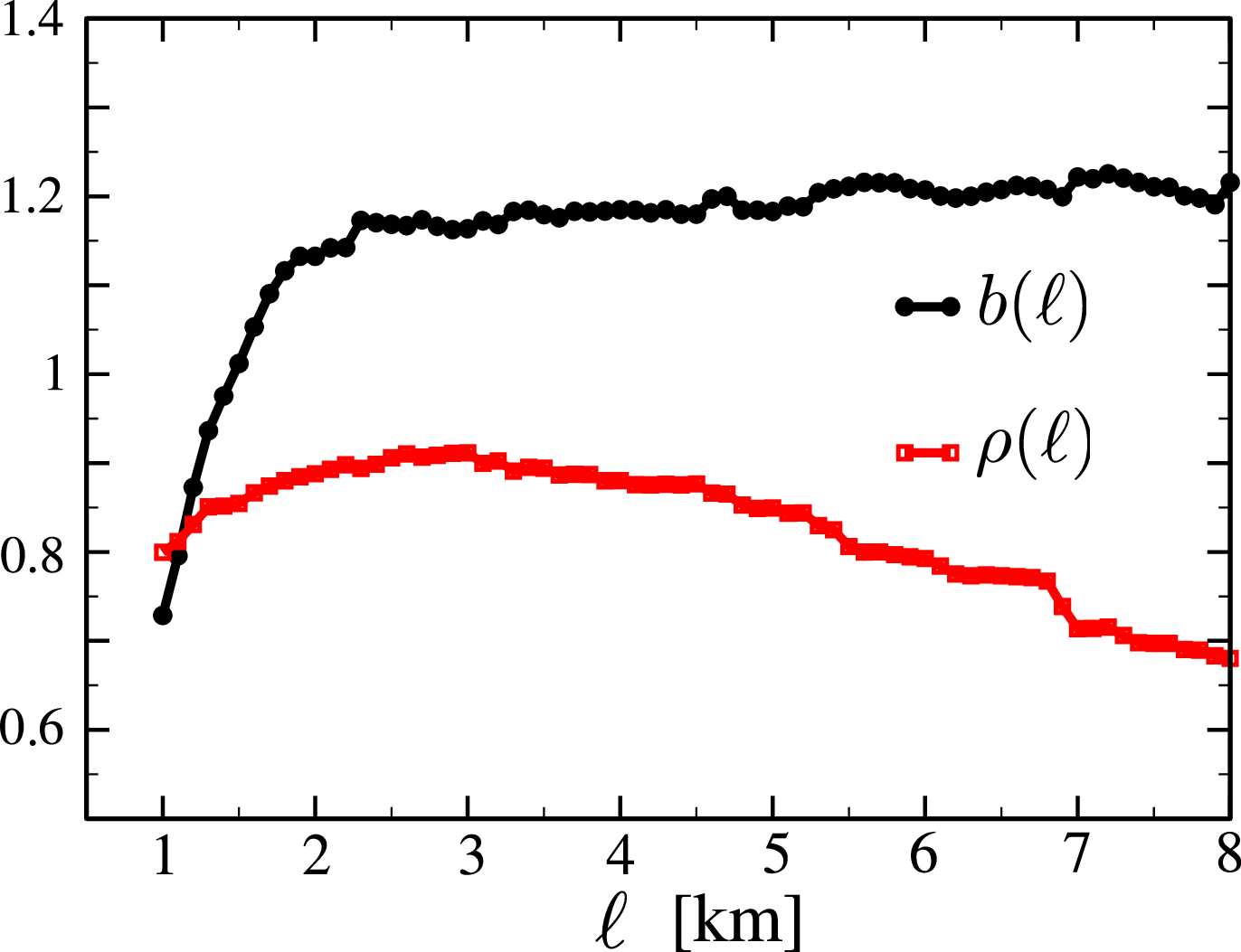}
}
\hbox{ {\bf c} \includegraphics[width=0.45\textwidth]{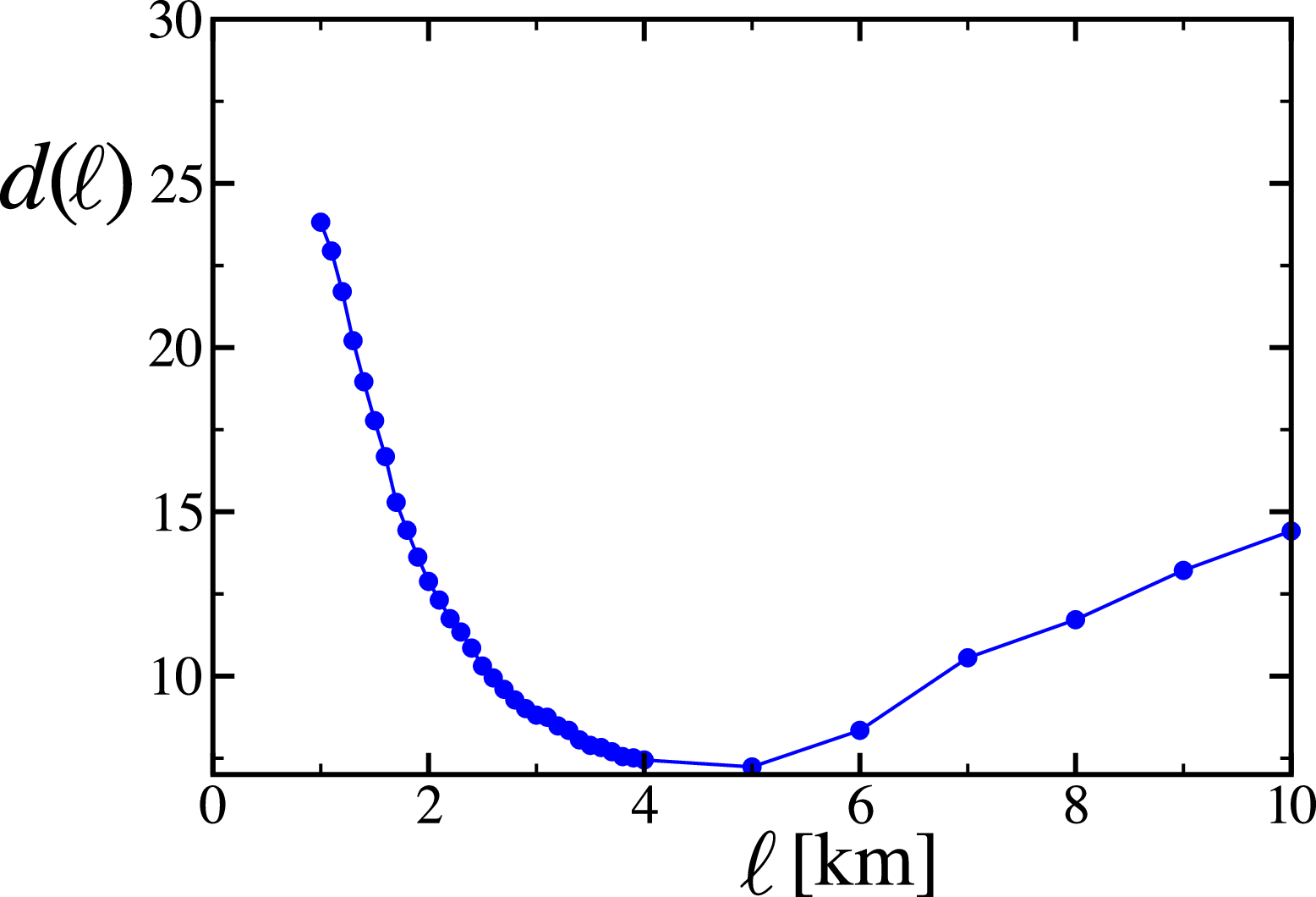}}
}
\caption{{\bf a,} Population of the CCA cluster in the USA for
$\ell = 3km$ vs its corresponding MSAs, using the one-to-one
correspondence explained in the text.
{\bf b,} Correlation analysis between CCA clusters and MSAs by plotting $\rho(\ell)$ for different values of $\ell$.
We quantify the
regression, $\ln S_{i}^{\rm CCA}\left( \ell\right) =a(\ell) +b(\ell)
\ln S_{i}^{\rm MSA}\left( \ell\right)$, by measuring the value of the
linear regression slope $b(\ell)$ as a function of $\ell$.
{\bf c,} Euclidean distance between MSAs and CCA clusters. }
\label{corr_CCA_MSA}
\end{figure}

In order to find the value of $\ell$ that best matches the MSAs we
match each MSA with the most populated overlapping CCA cluster. For
this purpose, from the US Census Bureau, we obtain the counties (and
corresponding FIPS) that belong to each MSA. An overlap between an
MSA and a CCA cluster exists if they share at least one FIPS code.
This overlapping procedure leads to several CCA clusters
corresponding to one particular MSA. To obtain a one-to-one
correspondence, among all overlapping CCA clusters we select the one
with the largest population.
We compare the
size of the obtained CCA cluster with the corresponding MSA by
computing the correlation, $\rho(\ell)$,
between the logarithm of the
cluster population, $S^{\rm CCA}_i(\ell)$, and the logarithm of the
population of the MSA, $S^{\rm MSA}_i$. Figure~\ref{corr_CCA_MSA}a
shows the cross-plot of $\log S_i^{\rm MSA}$ versus $\log S_i^{\rm
CCA}(\ell)$ for $\ell = 3$ km displaying an approximately linear behavior.
Figure~\ref{corr_CCA_MSA}b shows the correlation analysis
between CCA clusters and MSAs by plotting $\rho(\ell)$ for other
values of $\ell$. We quantify the regression, $\log S_{i}^{\rm
CCA}\left( \ell\right) =a(\ell) +b(\ell) \log S_{i}^{\rm MSA}\left(
\ell\right)$, by measuring the value of the linear regression slope
$b(\ell)$ as a function of $\ell$.
We find that $b(\ell)\approx 1.2$
for $\ell>2$ km.
Correlation in log sizes is very good for values of $\ell$ between 2
km and 6 km; the correlation, displayed in Fig.~\ref{corr_CCA_MSA}b, is very high
for this range of $\ell$. We find that $\rho(\ell)$ exhibits a
maximum value of $\rho\approx 0.91$ for $\ell\in[2.5,3.5]$ km, so
that we consider $\ell=3$ km as the optimal value.

We present another plausible measure of similarity between MSAs and
CCA clusters, based on the Euclidean distance. We define the
distance, $d(\ell)$, between MSAs and CCA as 
\begin{equation}
d(\ell) \equiv
\sqrt{\sum_i [{\ln}(S_i^{\rm MSA}) - {\ln}(S_i^{\rm
CCA}(\ell))]^2},
\end{equation}
where the sum is over all the MSAs and their
corresponding CCA clusters. In Fig.~\ref{corr_CCA_MSA}c we show the
distance between overlapped MSAs and CCA clusters as a function of
$\ell$. We find that when $\ell=5$ km the distance (in population)
is minimized, and that it is very low between $2.5$ km $\leq \ell \leq 6$ km in
approximate agreement with the log correlation analysis of Fig.~\ref{corr_CCA_MSA}a,b.

In addition to the MSAs, we compare the CCA clusters with US Census
Bureau ``places'' previously analyzed in~\citet{Eeckhout04} where a
log-normal distribution of city sizes was found. We first find a
one-to-one correspondence between CCA clusters and places, in
analogy to the previous match between MSAs and CCA clusters.  In
contrast to MSAs, US Census places take into account all towns,
villages, and cities and are based only on their administrative or
political boundaries~\citep{Eeckhout04,Holmes08}. The smallest and
largest places are Lost Spring, Wyoming, with exactly one resident,
and the political entity of New York City (Manhattan, Brooklyn,
Queens, Bronx, and Staten Island) with population 8.0 million.

From the US Census Bureau we obtain the geographical location of
each US Census place. Then, we identify each place with a unique
FIPS code. Accordingly, each place is associated with a unique CCA
cluster. This association leads to many places corresponding to a
single CCA. To obtain the one-to-one correspondence, among all
overlapping places we consider the one with the largest population.

\begin{figure}
\centering { \hbox{ {\bf a}
\includegraphics[width=0.4\textwidth]{pop_place_vs_cluster_place_2000_coarse=3000.eps}
{\bf b}
\includegraphics[width=.4\textwidth]{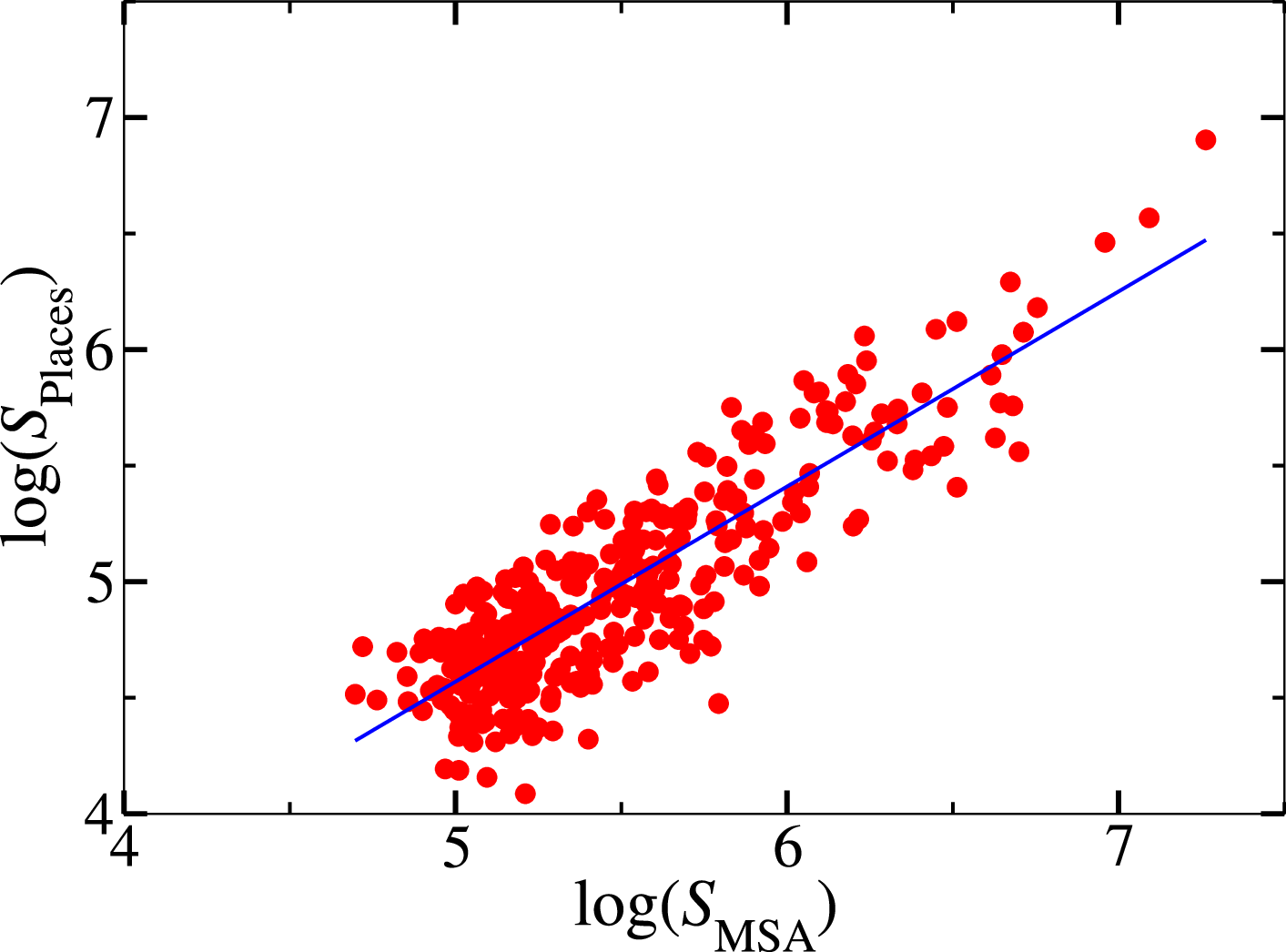}
}} \caption{{\bf a,} Log of population of US Census places vs. the
log of population of their corresponding CCA clusters for
$\ell=3$km. The straight line corresponds to a least square fit with
slope $b=0.90 \pm 0.02$ and y-intercept $a=0.25\pm0.09 $, from where
the correlation coefficient $\rho=0.79$ is obtained for cities with
population larger than 10,000. {\bf b,} Log of population of US
Census places vs. the log of population of their corresponding MSA.
The straight line corresponds to a least square fit with slope
$b=0.84 \pm 0.03$ and y-intercept $a=0.37 \pm 0.14$, from where the
correlation coefficient $\rho=0.87$ is obtained.}
\label{places_vs_CCA_MSA}
\end{figure}

In Fig.~\ref{places_vs_CCA_MSA}a we show that, the smallest cities found with
the CCA do not correspond well to US Census places; however, for cities above population $S=10,000$
CCA and Census places do exhibit a correlation coefficient of $\rho=0.79$. A detailed comparison
between CCA clusters and places shows that the number of small CCA
clusters is smaller than that for places because the CCA tends to
group small places that are geographically connected into a larger cluster. Therefore, the
construction based on places overestimates the number of small
cities and underestimates the number of large cities in comparison
with CCA, resulting in the size distribution of places to being less
fat tailed than the distribution for CCA clusters. This discrepancy,
which may find its root in the fact that places are purely based on
legal boundaries of locations~\citep{Holmes08}, may explain the
finding of a log-normal distribution of places~\citep{Eeckhout04},
whose full elucidation is beyond the scope of this paper.
Here, we show results for $\ell=3$ km as representative, but other
values of $\ell$ lead to the same conclusions.


We also perform a comparison between MSAs and places. In
Fig.~\ref{places_vs_CCA_MSA}b we observe a good congruence in the
whole range for which MSAs are defined. Notice that MSAs by
definition have a minimum population of 50,000. Therefore, when
looking for the one-to-one correspondence, only large places are
considered, leading to a good congruence, as found between large CCA
cluster and large places, with correlation
$\rho=0.87$.


\subsection{Robustness Checks}
\label{RobustnessCheck}

\begin{figure}[h]
\begin{centering}
\resizebox{0.6\textwidth}{!}{\includegraphics{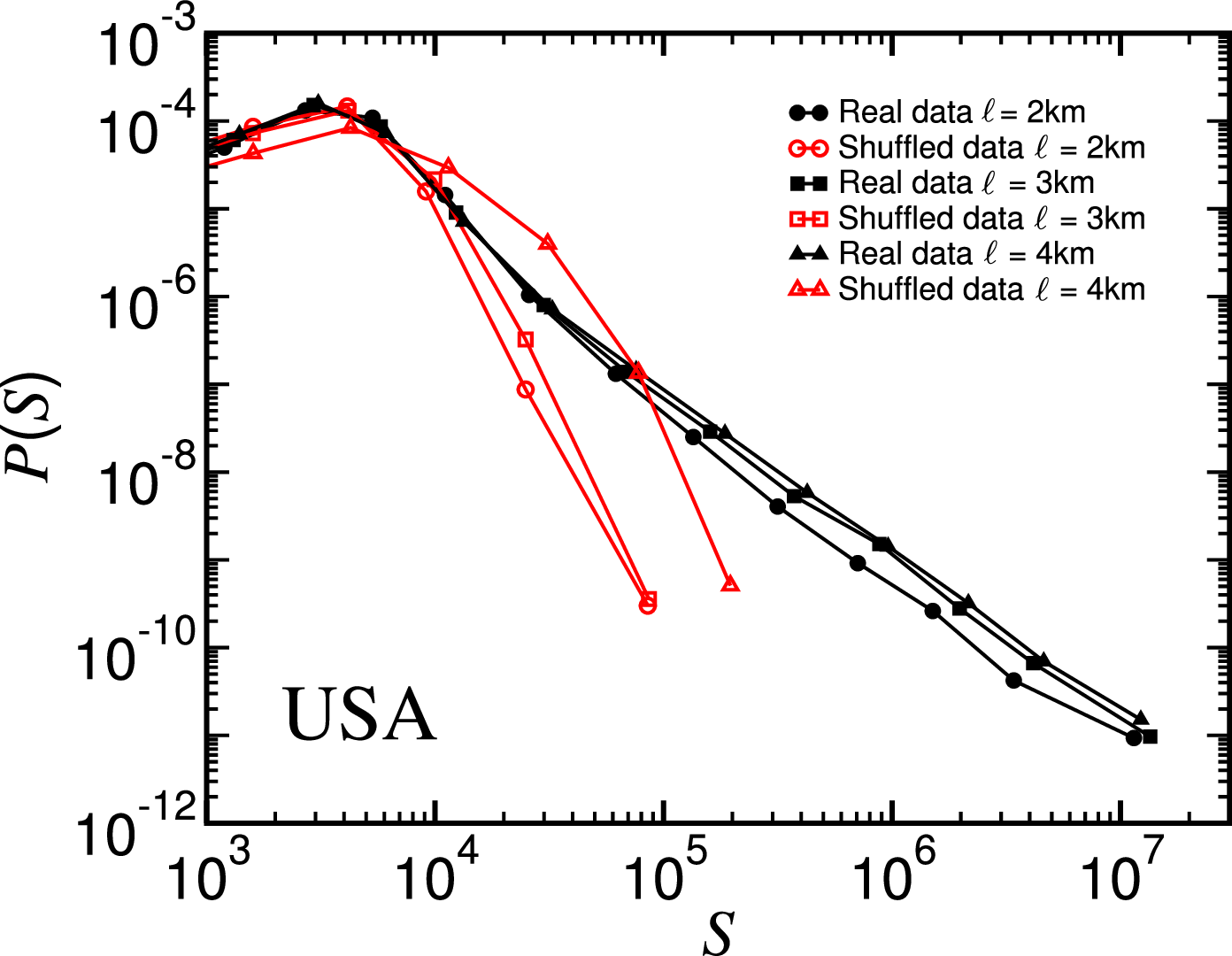}}
\caption{Population distribution for shuffled data. The black lines
correspond to the real data studied in Section~\ref{basicresults}. The red lines
correspond to the shuffled data, showing a change in the population
distribution and suggesting that the results of Section~\ref{basicresults}
are not an artifact of the CCA.} \label{shuffle}
\end{centering}
\end{figure}

In this section we test whether the results shown in
Section~\ref{basicresults} could be forced by the CCA, or
in other words, whether they could be an artifact of the CCA.
Starting with the actual location of the FIPS in the USA we
randomize the data by placing all 61,224 FIPS at random positions in
a rectangle of the same area as the USA. Then we apply the CCA to
obtain the corresponding clusters. This randomization procedure
preserves the population of each FIPS. In Fig.~\ref{shuffle} we show
the population distribution for the shuffled data and for the
original data. These results show that the shuffled data does not
exhibit Zipf's law. The largest cluster for the shuffled data
contains 196,112 inhabitants: the reshuffling prevents the emergence
of very large clusters. This suggests that the CCA is not forcing
the data to present a power-law for the population distribution, and
that Zipf's law arises purely from the data.

\section{Investigation of the Geography of Cities: Areas and Densities}
\label{area}

\subsection{Areas}

The CCA presents a unique feature in that it allows the definition of the area of cities not based on administrative boundaries. Such a feature in not present in agglomerations defined by Places or MSAs. 
Thus, the spatial analysis of the CCA allows us to examine a possible feature
of the origin of the Zipf's law: highly populated cities may have a
large geographic area. Therefore, it is of interest to study the
distribution of areas~\citep{Makse95}, $P(A)$, defined by the CCA.

As explained above, the data of GB consists of a high resolution grid with cell size $200$m. 
Therefore, after applying the CCA, we calculate the area of a cluster in GB as the 
number of cells in the cluster multiplied by the area
of a cell, $\ell^2$.

The case of the USA is more complicated. The data consists of 61,224 populated points
on the map. Each point corresponds to a different FIPS code,
defined by the US Census Bureau.  USA FIPS are simply a partition of
the map of the USA, so that any point in the map belongs to one FIPS
code, and each FIPS has an associated area which is given by the US
Census Bureau in the dataset.
In the USA, FIPS codes are not homogeneously distributed. In the New
York City area, there is high resolution, which means that there are
many FIPS covering a small area, but in the state of Wyoming or Utah
the resolution is quite low, so that there are FIPS with a large
area. For instance, FIPS in Manhattan typically cover an area of about
0.20 km$^2$ while in the state of Utah FIPS 49003960100 covers a large
area of 15,962 km$^2$. Therefore, when $\ell$ is of the order of a
few kilometers, a FIPS in the Wyoming
area will remain isolated in its own cluster, but still its area will
be extremely large, typically a couple of orders of magnitude larger
than $\ell^2$. Therefore, since the area of isolated points is very
large, these points will appear at the tail and in the middle of the
distribution $P(A)$, overestimating the outcome for middle and large
areas. Accordingly, in order to compute the $P(A)$, we do not take into
account clusters containing only 1 or 2 FIPS since they
overestimate the amount of land they cover. Moreover, the population
of those isolated points is typically small and rarely exceeds $S=10,000$. In fact, 
we find that removing all clusters with only 1 or 2 FIPS
is practically the same as removing all clusters with population
smaller than 10,000: only 7\% of clusters with 1 or 2 FIPS have a
population larger than 10,000.

\begin{figure}
\centering { \hbox{
{\bf a}
\includegraphics[width=0.45\textwidth]{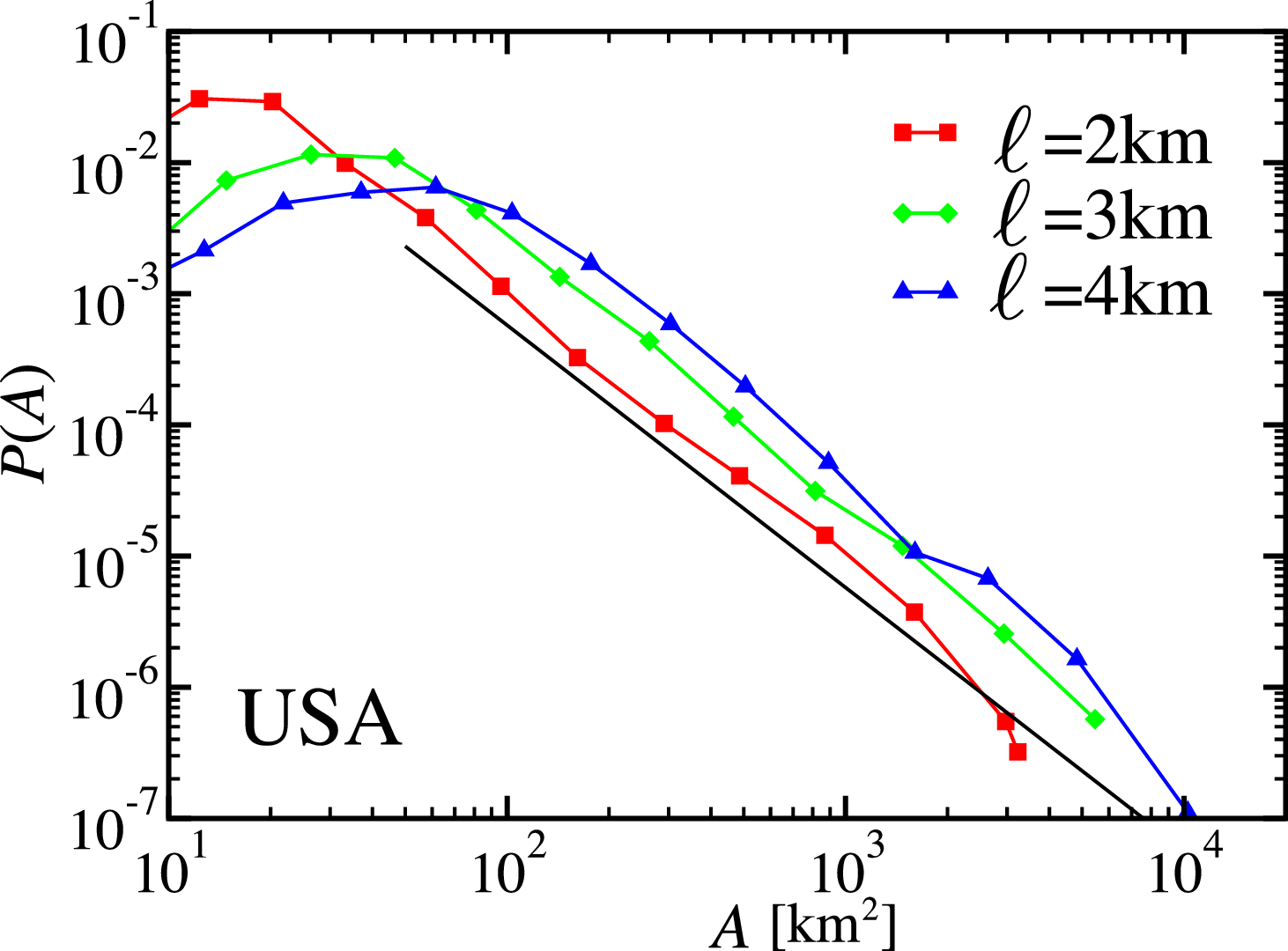}
{\bf b}
\includegraphics[width=.45\textwidth]{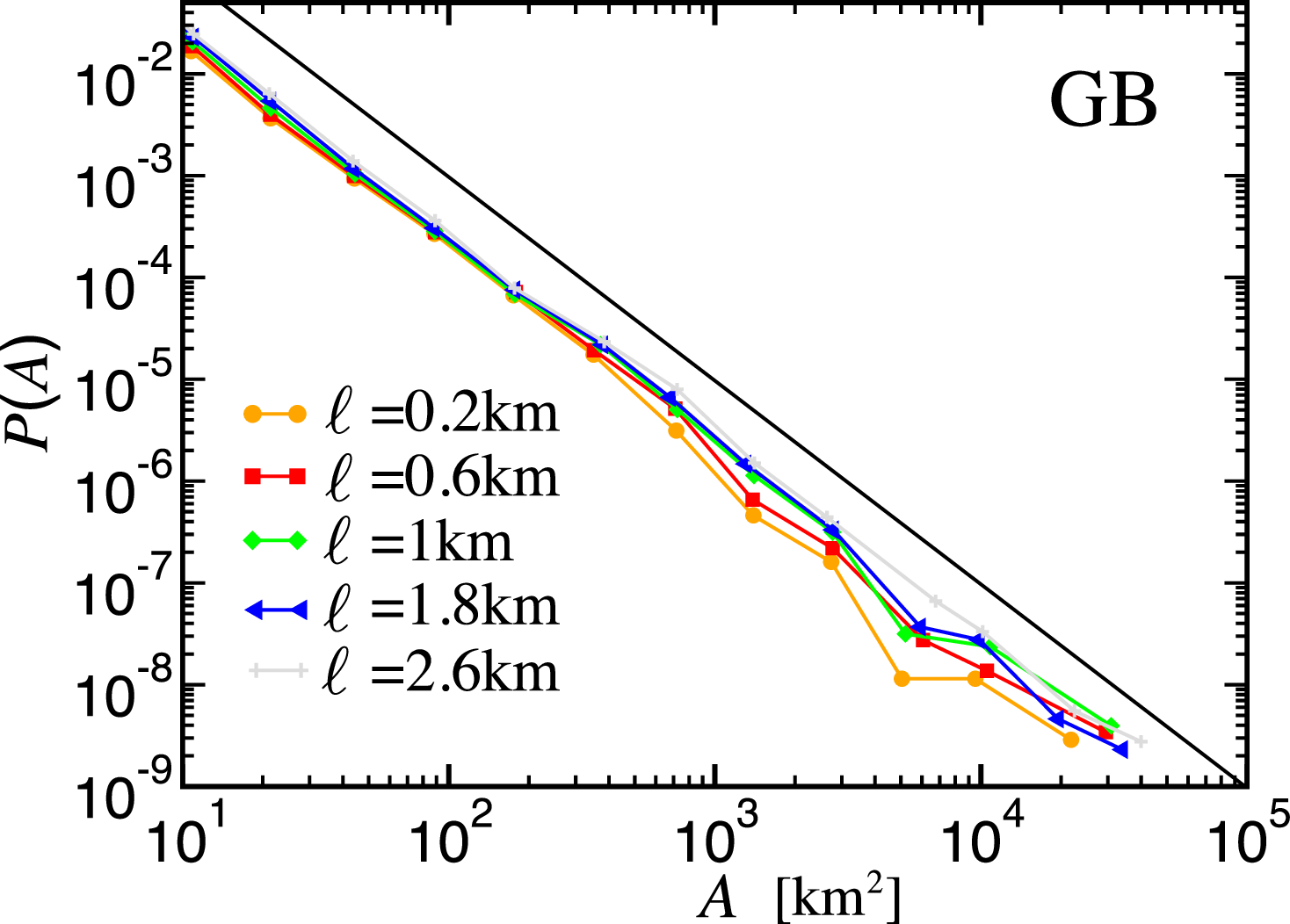}
}}
\caption{{\bf a,}
Probability
distribution of the areas, $P(A)$, for the USA for different
$\ell$.
{\bf b,}
Probability distribution $P(A)$ of the areas of the
clusters in GB at different coarse-graining scales $\ell$.  The
distribution of city areas for GB is also consistent with Zipf's
law. We find $\zeta_{A}=0.97\pm 0.04$, for $\ell=1$ km.  The black
solid lines denote Zipf's law, i.e. a power-law function with exponent
-2.
}
\label{areas}
\end{figure}

In Fig.~\ref{areas}a we report the results of $P(A)$ for the USA. We
find a power-law distribution of the form 
\begin{equation}
P(A)\sim A^{-\zeta_{A}-1},
\end{equation}
 with a Zipf exponent $\zeta_{A}=1.07\pm 0.04$, for $\ell=3$ km.  
In
Fig.~\ref{areas}b we show the results of $P(A)$ for GB. As for the
USA, we find that the area distribution for GB follows a power-law
with exponent $\zeta_{A}=0.97\pm 0.04$, for $\ell=1$ km.
This extends the results obtained in~\citep{Makse98} for areas distributions surrounding a city like London and  Berlin~\citep{Makse95} and in UK~\citep{Makse98}. The result of the Zipf's law for areas in the US appears to be new.

This result may be an important update for calibrated
models of cities where transport costs of goods or people play an
important role~\citep{brakman01,Fujita01}. The Zipf's law for areas
implies that some cities have very large areas, and those cities' viability
may mean that transport costs cannot be
too large, or are mitigated in economically interesting ways. We come back to
this topic in Section~\ref{model}.

\begin{figure}
\centering { \hbox{ {\bf a}
\includegraphics[width=0.45\textwidth]{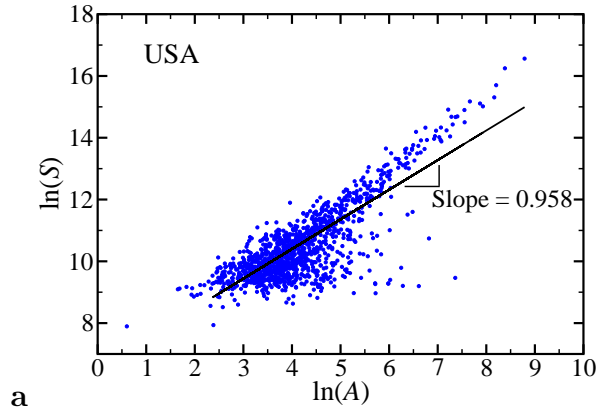}
}}
\caption{Logarithm of the population, $S$ versus the logarithm of
the area, $A$, for {\bf a,} the USA with $\ell=3$ km and {\bf b,} GB
for $\ell=1$ km. The black lines denote the OLS regression (see
Table~\ref{area_pop_table}.)} \label{corr_area_pop}
\end{figure}

\begin{table}[t]
\caption{Results of the OLS regression analysis of ${\rm ln}S = a +
b~{\rm ln}A$, where $A$ is the area and $S$ the population. We
report results for $S_*=12,000$ and $ \ell=3$ km for the USA, and
$S_*=5,000$ and $\ell=1$ km for GB. Standard errors are reported in
parentheses.} \label{area_pop_table}
\begin{center}
\begin{tabular}{ccc}
\hline \hline
& USA & GB \\
\hline
ln$A$ & 0.958 & 1.065\\
& (0.020) & (0.007)\\
Constant & 6.567 & 8.166\\
& (0.085) & (0.010)\\
Observations & 1064 & 1007\\
$R^2$ & 0.686 & 0.921 \\
\hline
\end{tabular}
\end{center}
\end{table}

In Fig.~\ref{corr_area_pop}a we study the correlations between areas
and populations for the USA and GB. We find that the linear OLS
regression ${\rm ln}A = a + b~{\rm ln}S$ leads to the results shown
in Table~ \ref{area_pop_table}, indicating a strong correlation
between areas and population in log sizes. Indeed, the finding of
$b\simeq1$ indicates that population is, to a good degree of
approximation, simply proportional to area. This finding motivates
us to study city density in more detail.



\subsection{Densities}

In this section we study the population density, $D=S/A$.
\footnote{See \citet{bryan07} for an alternative
analysis of density. They find that density has fallen in the US
over the past seven decades.}
We study the behavior of $D$ versus $S$ and  $A$ by performing
the linear regressions ${\rm ln}D = a + b~{\rm ln}A$, and ${\rm ln}D
= a + b~{\rm ln}S$. Table~\ref{DvsSandDvsA} shows the results of the
OLS regression estimates with $S_*=12,000$ and $\ell=3$ km for the
USA, and $S_*=5,000$ and $\ell=1$ km for GB (other choices of $\ell$
lead to the same conclusions). We find that population density has
very little relation to the area: the coefficients are very close to
0. It has a slightly higher link with population. Of course,
measurement error in the variables may bias the measurement.

Still, the link between density and area is perhaps
surprisingly weak. Some urban systems, like New York City, are quite dense,
but even then, the effects are moderate: the density of New York City is only 3.7 
times the national median even though its population is 485 times the national 
median. Of course, we obviate here a consideration of the
interesting heterogeneity within cities; but for the purposes of this paper such a study may be
deferred to later work.
We find that density has a very small dispersion: the standard deviation
of its natural logarithm is 0.28 for the USA and 0.09 for GB. In contrast,
the corresponding quantity for areas and population is about 1.
Hence, we conclude that city area covaries greatly with
population, and little with density. We next propose a model that is
consistent with this finding, as well as the power law scaling of
city sizes.


\begin{table}[t]
\caption{Results of the OLS regression analysis of ${\rm ln}D = a +
b~{\rm ln}A$ and ${\rm ln}D = a + b~{\rm ln}S$, where $D=S/A$ is the
density, $A$ the area, and $S$ the population. We report results for
$S_*=12,000$ and $\ell=3$ km for the USA, and $S_*=5,000$ and
$\ell=1$ km for GB. Standard errors are reported in parentheses.}
\label{DvsSandDvsA}
\begin{center}
\begin{tabular}{cccccc}
\hline \hline
& \multicolumn{2}{c}{${\rm ln}D = a + b~{\rm ln}A$} & ~~~ & \multicolumn{2}{c}{${\rm ln}D = a + b~{\rm ln}S$}\\
\cline{2-3} \cline{5-6}
& USA & GB & & USA & GB \\
\hline
ln$A$ & -0.042 & 0.065 & ln$S$ & 0.284 & 0.099\\
& (0.020) & (0.007) & & (0.015) & (0.006)\\
Constant & 6.567 & 8.166 & Constant & 3.357 & 7.299\\
& (0.086) &  (0.010) & & (0.159) & (0.057)\\
Observations & 1064 & 1007 & Observations & 1064 & 1007\\
$R^2$ & 0.004 & 0.007 & $R^2$ & 0.256 & 0.050\\
\hline
\end{tabular}
\end{center}
\end{table}

\section{Model}
\label{model}

Recent economic theories that are compatible with Zipf's law generally
rely on the existence of random growth~\citep{Champernowne53,Simon55,
Krugman96,Levy96,Gabaix99,Dobins00,davis02,GabaixIoannides04,
Eeckhout04,Duranton06,Duranton07,RossiWright07,Cordoba08}:
cities follow a proportional growth
process where the distribution of the percentage growth rate is the same
for small and large
cities. Small cities, however, grow faster~\citep{Glaeser92,Glaeser95,RozenfeldPNAS}, 
which prevents the distribution
from becoming degenerate. Some theories obtain Zipf's law only approximately, and do not
obtain it over the range that we find in the present work. Accordingly, we present a
parsimonious model that
generates an approximate Zipf's law for population and area.

We first describe the model at a given point in time. Cities are
indexed by $i\in \left[ 0,1\right] $. City $i$ employs $S_{i}$
workers, and has a competitive sector producing good $i$, which it produces in quantity $%
y_{i}=b_{i}S_{i}$, where $b_{i}$ is the productivity. The aggregate
good is a Dixit-Stiglitz aggregator with elasticity of
substitution $\eta >1$:%
\begin{equation}
Y=\left( \int y_{i}^{\left( \eta -1\right)/\eta}di\right) ^{\eta
/\left( \eta -1\right) }
\end{equation}

There is a potentially unbounded quantity of land, but making usable
an area $a$ of land necessitates an investment $p a$, for some unit
cost $p$. This reflects e.g. maintenance cost, roads and other
infrastructure to occupy a land area $a$ (the cost is in the
consumption good, but it could equivalently be in units of labor).
Hence, as in \citet{RossiWright07}, land use is endogenous. As a
result, if $A$ is total land use, and $C$ total consumption, the
resource constraint is $C+pA\leq Y$.

Consumers' utility is $u\left( c,a\right) =c^{1-\beta }a^{\beta }$ with $%
0<\beta \,<1$, where $c$ is the consumption of the good, and $a$ the
consumption of land. Workers are free to choose their cities, so that utilities
are the same across cities. Hence, the competitive
equilibrium is also the solution to the planner's problem that
equalizes utility across agents, and allocates population $S_{i}$ in
each city $i$ (subject to $\int S_{i}di=S$, the total population),
and allocates their per capita consumption of good $c_{i}$ and land
area $a_{i}$, to maximize total utility subject to the resource constraint:%
\[
\max_{S_{i},c_{i},a_{i}}\int u\left( c_{i},a_{i}\right)
S_{i}di\text{ subject to}
\]%
\begin{eqnarray*}
\int \left( c_{i}+pa_{i}\right) S_{i}di &\leq &\left( \int \left(
b_{i}S_{i}\right) ^{\left( \eta -1\right) /\eta }di\right) ^{\eta
/\left(
\eta -1\right) } \\
\forall i,j,u\left( c_{i},a_{i}\right)  &=&u\left( c_{j},a_{j}\right)  \\
\int S_{i}di &=&S
\end{eqnarray*}

The solution method is standard. The labor allocated to producing good $i$, i.e., the population living in city $i$ is:%
\begin{equation}
S_{i}=\frac{B_{i}}{\overline{B}}S  \label{L}
\end{equation}%
where $B_{i}\equiv b_{i}^{\eta -1}$ and $\overline{B}\equiv \int
B_{i} di$. GDP is $Y=\overline{B}^{1/\left(\eta -1\right)}S$. A
fraction $1-\beta $ of income is devoted to consumption of the good,
and a fraction $\beta$ to land use. Each consumer purchases a
quantity of land $\nu \equiv \beta \overline{B}^{1/\left(\eta
-1\right)}/p$. So, the total quantity of land in city $i$, $A_i$, is $\nu$
times the number of inhabitants of city $i$:
\begin{equation}
A_{i}=\nu S_{i}  \label{A}
\end{equation}%
Hence city area and city population are proportional.

Next, we wish to see why Zipf's law might arise.
We consider a
dynamic version of the above static description.
Consumers have utility %
$E \left[\int e^{-\delta t} u(c_{t},a_{t}) dt \right]$ with some
discount rate $\delta$, which is assumed to be sufficiently large
for utility to be finite. As there are no adjustment costs, for
given productivities, the dynamic model yields the same allocation
of workers and land across cities as in the static model. We take a
model that merges the
random growth models of cities and the model of random growth of firms
developed by~\citet{Luttmer07}. We
postulate that (elasticity-adjusted) productivity $B_i$ of city $i$
evolves as a geometric Brownian motion:%
\begin{equation}
\frac{dB_{it}}{B_{it}}=gdt+\sigma dz_{it} \label{dB}
\end{equation}%
where $g$ is the mean of the growth rate of productivity of an
existing city, $\sigma$ the volatility of that growth rate, and
$z_{it}$ are independent Brownian motions. However, if a city is too
unproductive, it can \textquotedblleft refresh\textquotedblright \
its
productivity as a fraction of the average productivity:%
\begin{equation}
B_{it}\geq \pi \overline{B}_{t} \label{B barrier}
\end{equation}%
where $\pi \in \left( 0,1\right) $ is a constant. Here, we simply
postulate that it can reset its productivity for free, by simply
imitating the average productivity, but only imperfectly:\ its reset
productivity is only a fraction $\pi $ of the average productivity.
\citet{Luttmer07} presents a much more elaborate microfoundation for
this idea, including the $\pi $, but the above model is useful for
its simplicity. All in all, $B_{it}$ follows a geometric Brownian
motion, reflected at $\pi \overline{B}_{t}$.

The following Proposition characterizes the behavior of this
economy.

\medskip
\noindent {\bf Proposition }{\it (i)\ The steady state distribution
of city population and city area is a power-law with exponent $\zeta
$:\begin{equation} \zeta =\frac{1}{1-\pi }
\end{equation}%
Indeed, \ $S_{it}/\overline{S}_{t}$, $A_{it}/\overline{A}_{t}$ and
$B_{it}/\overline{B}_{t}$ are all equal and follow the Pareto
distribution $P\left( X\geq x\right) =\left(x/\pi \right) ^{-\zeta
}$ for $x\geq \pi $.  The exponent $\zeta $ tends to 1 (the Zipf's
law value) when the friction $\pi $ coming from the reflecting
barrier tends to 0.

(ii)\ City population $S$ is proportional to city area $A$, and
density $D=S/A$ is independent of city size.

(iii)\ The fraction of income spent on housing is independent of
city size.}

\medskip
\noindent \textbf{Proof} The proof method is as in \citet{Gabaix99}
(see also \citet{Gabaix09} and the references therein). Denote by
$\overline{g}$ the growth rate $\overline{B}_{t}$ on the balanced
growth path. The relative share of city $i$, $s_{it}\equiv
B_{it}/\overline{B}_{t}$ follows a geometric Brownian motion, with
$ds_{it}/s_{it}=(g-\overline{g})dt+\sigma dz_{it}$, with a
reflecting barrier, $s_{it}\geq \pi $. Calling $\mu
=g-\overline{g}$, the steady state density $p\left( s\right) $
follows the Forward Kolmogorov
equation:%
\[
0=-\left( \mu sp\left( s\right) \right) ^{\prime }+\frac{1}{2}\left(
\sigma ^{2}s^{2}p\left( s\right) \right) ^{\prime \prime }
\]%
Integration of this equation yields $p\left( s\right) =ks^{-\zeta -1}$ for $%
\zeta =1-2\mu /\sigma ^{2}$ and some constant $k$. By construction $E\left[ s%
\right] =1$. Given $\int_{\pi }^{\infty }p\left( s\right) ds=1$, we
have
\[
1=\frac{\int_{\pi }^{\infty }p\left( s\right) sds}{\int_{\pi
}^{\infty }p\left( s\right) ds}=\frac{\int_{\pi }^{\infty
}ks^{-\zeta-1 }sds}{\int_{\pi }^{\infty }ks^{-\zeta -1}sds}=\pi
\frac{\zeta }{\zeta -1}
\]%
which yields $\zeta =1/\left( 1-\pi \right) $. The steady state
distribution
can be written $P\left( s_{it}\geq x\right) =\left( x/\pi \right) ^{-\zeta }$%
. Finally, by (\ref{L}) and (\ref{A}), the distribution of
populations and areas is a Pareto with the same exponent $\zeta $.

We also note that $\zeta =1-2(g-\overline{g})/\sigma ^{2}$. This
yields the value of the growth rate of productivity:
$\overline{g}=g+\sigma ^{2}\left( \zeta -1\right) /2$. The
(endogenous)\ growth rate of average productivity is higher than the
(exogenous)\ growth rate of a city above the reflecting barrier,
because this reflecting barrier makes small cities grow faster. In
the Zipf limit where $\pi \rightarrow 0$, hence $\zeta \rightarrow
1$, the difference between the two growth rates, $\overline{g}-g$,
goes to 0. $ \blacksquare $
\medskip

This economy reflects our main empirical findings, (i) and (ii).
Point (iii)\ reflects the findings of \citet{davis08}, who find
that the fraction of income spent on housing is roughly constant
over time and across city sizes.

We note that here, following \citet{RossiWright07} and \citet{stijn09}, land is not
exogenous but instead it is acquired. This is a legitimate modelling
idealization in our view. Take a city such as Dallas, which starts
with vast quantities of unoccupied land around it. It can grow in a
fairly unlimited way, but it needs to pay for the land use, e.g.
building infrastructure such as road, electricity and running water.
It makes sense to model this activity as a constant-return to scale
activity, at least in the first approximation. At the other end of
the spectrum, we may have New York. But even it has grown
considerably by geographical expansion, which lends credence to our
model. It would be interesting, and surely desirable, to extend the
model with some sort of increasing cost of land use (given some
limit). We conjecture that, if the random growth effects are large
enough, this will modify the power law distribution, but will not
eliminate it. A calibration of the deviation from the constant
return to scale model, and the deviation of the power law, would be
useful, but we will not attempt it here.

Here cities are basically constant-return-to-scales economies,
except for one large Marshallian force that makes a given good only
producible in one city (as ``secrets of the trade'' may be exclusive
that city). Of course, this is a stark model, but it is
parsimonious, and is consistent with our scaling facts. In addition,
external effects linked to cities may not be huge. For
instance,~\citet{glaeser98} reports quantitatively moderate
deviations from the hypothesis that cities are constant-return-to
scale (see also ~\citet{Bettencourt07}). For instance,
~\citet{glaeser98} reports that the average commute time in cities
of less than 100,000 is 20.5 minutes each way, while in cities of
more than 1,000,000 it is 31.9 minutes each way\footnote{In a related vein,~\citet{Ciccone96} estimate that a doubling of density increases productivity by 5.5\%, while~\citet{Davis09} finds an increase of 2\%. This is a arguably small deviation from the constant-return to scale benchmark we use in our model.}.
This difference may
be small compared to the huge differences in size and area that our
model focuses on.

Our model postulates that Gibrat's law holds. However, deviations from Gibrat's law have  been found in the literature and for the CCA clusters (e.g. \citet{Glaeser95}, and \citet{RozenfeldPNAS}). A simple theoretical solution to this apparent tension between the data and the idealization used in models based on Gibrat's law is discussed in \citet{GabaixIoannides04}, Section 3.2.2. Urban growth may accommodate a wide range of growth processes exhibiting a Pareto distribution, but also deviations from Gibrat's law, as long as
they contain a unit root (which satisfies Gibrat) with respect to the logarithm of city size: in particular, growth
processes can have some mean-reverting component that violates Gibrat's law. Under that hypothesis, the deviations from Gibrat's law would come from the mean-reverting component of the growth process, but Gibrat's law in the unit root part of the process would ensure the Pareto law. More research is needed to empirically assess this possibility. It would likely require empirical studies of Gibrat's law over long time intervals.

We think that the model could be extended to add positive and
negative agglomeration externalities, which also can generate random
growth, as in~\citet{Gabaix99aer},~\citet{Eeckhout04}, and
~\citet{RossiWright07}. We also eschew a detailed modelling of the
heterogeneity within a city, such as the one in \citet{lucas02}.
Such developments would be very welcome, but we propose to defer
them to future research.

\section{Conclusion}
\label{conclusions}

We have used a ``bottom-up'' approach which allows us to construct
cities independently of their ``legal'' definition, instead using
a more geographical and economic basis. The resulting data extend the domain
of validity of Zipf's
law to a considerable range: we show that when cities are
constructed independently of their administrative boundaries, Zipf's
law appears to be a genuine regularity for the bulk of the city size
distribution. Second, we are able to analyze city areas, which
allows for the estimation of a potentially very important quantity
in urban economics, and anchors the definition of cities much more in geography.
We find evidence for a power-law
distribution of areas, with an exponent close to 1. Third, we
presented a model incorporating both population and area, that
matches our \textquotedblleft macro\textquotedblright \ facts.
Fourth, we provide a public good by putting on our web page the
correspondence between ZIP code and our Clusters, so that other
researchers can use the agglomerations constructed with the CCA, and
study dimensions of local economics other than areas and
populations.

In the present work we have investigated only two countries. It is
natural to extend this study to more countries, an investigation
that might offer confirmation of the scaling laws for areas,
population and density that we have found, and also perhaps find
economically interesting deviations from them.
The minimalist model presented here could be extended to incorporate
richer specification of the internal structure of cities. We think
that this \textquotedblleft bottom-up\textquotedblright \ approach
could be useful for a host of urban questions. Combining our
geographical approach with land price data could lead to a much
more constrained and geography-based theory of the macro and internal structure of cities.


\clearpage

\bibliography{refs}

\end{document}